\documentclass[onefignum,onetabnum]{siamart190516}
\include{./anc/commands}
\input{./anc/shared}

\ifpdf
\hypersetup{
  pdftitle={Compute-bound Galerkin ROM of LTI systems},
  pdfauthor={F.Rizzi, E.J.Parish, P.J.Blonigan, and J.Tencer}
}
\fi

\begin{document}

\maketitle

% REQUIRED
\begin{abstract}
  % what is this about and why?
  This work aims to advance computational methods for
  projection-based reduced-order models (ROMs) of
  linear time-invariant (LTI) dynamical systems.
  For such systems, current practice relies on ROM formulations
  expressing the state as a rank-1 tensor (i.e., a vector),
  leading to computational kernels that are memory bandwidth bound
  and, therefore, ill-suited for scalable performance
  on modern architectures.
  This weakness can be particularly limiting when
  tackling many-query studies, where one needs to run a large number of simulations.
  % articles describes what?
  This work introduces a reformulation, called rank-2 Galerkin,
  of the Galerkin ROM for LTI dynamical systems which converts
  the nature of the ROM problem from memory bandwidth to compute bound.
  %, making it scalable and efficient on modern computing node architectures.
  We present the details of the formulation and its implementation, and
  demonstrate its utility through numerical experiments using, as a test case,
  the simulation of elastic seismic shear waves in an axisymmetric domain.
  We quantify and analyze performance and scaling results for varying numbers
  of threads and problem sizes.
  Finally, we present an end-to-end demonstration of using the rank-2 Galerkin
  ROM for a Monte Carlo sampling study. We show that the rank-2 Galerkin ROM
  is one order of magnitude more efficient than the rank-1 Galerkin ROM
  (the current practice) and about $970$ times more efficient
  than the full-order model, while maintaining accuracy
  in both the mean and statistics of the field.
\end{abstract}

% REQUIRED

% REQUIRED

%-------------------------------------------------------
%                       INTRO
%-------------------------------------------------------

\section{Introduction} \label{sec:intro}
The ``extreme-scale'' computing era in which we are living, 
also referred to by many as the dawn of exascale computing (the capability 
to perform $10^{18}$ floating-point operations per second), 
is enabling a shift in the paradigm within which we approach new problems.
We no longer target few simulations executed for specific
choices of parameters and conditions, but are increasingly
interested in exploiting high-fidelity models to generate
ensembles of runs to tackle {\it many-query} problems,
such as uncertainty quantification (UQ).
These typically require large sets of runs to adequately
characterize the effect of uncertainties in parameters and operating
conditions on the system response.
Such an approach is key to solve, e.g., design optimization
and parameter-space exploration problems.

If the system of interest is computationally expensive
to query---for example very high-fidelity models for which
a single run can consume days or weeks on a supercomputer---its use
for many-query problems is impractical or even impossible.
Consequently, analysts often turn to surrogate models, which replace the
high-fidelity model with a lower-cost, lower-fidelity counterpart.
To be useful, surrogate models should meet
accuracy, speed, and certification requirements.
Accuracy ensures that the surrogate produces a sufficiently small
error in target quantities of interest. The maximum acceptable
error is typically defined by the user and is problem-dependent.
Speed ensures that the surrogate evaluates much more rapidly
than the full-order model, and one typically has to consider a
tradeoff between speed and accuracy.
Certification ensures that the error (and its bounds) introduced
by the surrogate can be properly quantified and characterized. 
We additionally note that it is often desirable that surrogate 
models preserve important physical properties, 
such as Lagrangian or Hamiltonian structure, or mass conservation.

Broadly speaking, surrogate models fall under three categories,
namely (a) \textit{data fits}, which construct an explicit
mapping (e.g., using polynomials, Gaussian processes)
from the system's parameters (i.e., inputs) to
the system response of interest (i.e., outputs), (b) \textit{lower-fidelity
models}, which simplify the high-fidelity model (e.g., by coarsening the mesh,
employing a lower finite-element order, or neglecting physics), and (c)
\textit{projection-based reduced-order models (ROMs)}, which reduce the number
of degrees of freedom in the high-fidelity model though a projection process.
The main advantage of ROMs is that they apply a projection process
directly to the equations governing the high-fidelity model,
thus enabling stronger guarantees (e.g., of structure preservation, of
accuracy via adaptivity) and more robust error analyses, e.g.,
via \textit{a priori} or \textit{a posteriori} error bounds.
%accurate \textit{a
%posteriori} error analysis (e.g., via \textit{a posteriori} error bounds or
%error models).
We can identify two main research branches within the field of ROMs,
one addressing nonlinear systems and the other focusing on linear systems.
In this work, we focus on the latter, more specifically
linear time-invariant (LTI) dynamical systems, which are
linear in state but have an arbitrary nonlinear parametric dependence.

Projection-based model reduction of LTI systems
is a mature field in terms of methodological
and algorithmic development \cite{BeGuWi15,BaBeFe14},
but we argue that it lags behind in terms of implementation
and computational advancement.
On the one hand, numerous ROM techniques have been developed 
for such systems accounting 
for, e.g., observability and controllability~\cite{Mo81,willcox2002bmr,Ro76, LALL19992598},
$\mathcal{H}_2$-optimality~\cite{GuAnBe08,Wi70,HyBe85},
structure preservation~\cite{LALL2003304,BeSaUd16},
and non-affine parametric dependence~\cite{BuWiGh08,GrPa05}. Furthermore, these ROMs can be \textit{certified} via \textit{a priori} and \textit{a posteriori}
error analysis~\cite{reliable_prediction_rb,rovas_apost,Ro03,KuVo01,Si14, Volkwein12modelreduction,Mo81,BeGuWi15,GuAt04}.
As a result, it is possible to construct stable,
accurate, and certified ROMs for a wide class of LTI systems.
On the other hand, the computational
aspects of solving these systems efficiently, especially in the context
of many-query problems, has not achieved a comparable level of maturity.
This statement is grounded in the fact that the standard formulation
of ROMs for LTI dynamical systems expresses
the state as a rank-1 tensor (i.e., a vector), see \cite{BeGuWi15,BaBeFe14},
which makes the corresponding computational kernels memory bandwidth bound
(as we will show in \S~\ref{sec:romScaling}), and therefore not well-suited
for modern multi-core processors or accelerators.

This work presents our contribution towards improving
the computational efficiency of ROMs for LTI systems.
We present and demonstrate a reformulation
of the ROM problem for LTI dynamical systems such that we change
its nature from memory bandwidth bound to compute bound, making it
more suitable for modern multi- and many-core computing nodes.
More specifically, we believe this work makes the following contributions:
\begin{itemize}
\item a new ROM formulation, referred to as ``rank-2 Galerkin'',
  for LTI dynamical systems that is efficient for many-query problems,
  and comprehensive numerical results to demonstrate its performance and scalability;
\item a new open-source C++ code, developed from the ground up using
the performance-portable Kokkos programming model \cite{CarterEdwards:2014},
to simulate the evolution of seismic shear waves in
an axi-symmetric domain;
\item detailed numerical examples based on the shear wave problem
to demonstrate the rank-2 Galerkin ROM (which, to
the best of our knowledge, also constitutes the first application
of ROMs to seismic shear waves).
\end{itemize}

The paper is organized as follows.
In \S~\ref{sec:formulation}, we present the formulation and discuss
the advantages and disadvantages of various Galerkin ROM formulations,
in \S~\ref{sec:testcase} we present
the test case chosen for our numerical experiments and its implementation details,
and in \S~\ref{sec:results} we describe the results.
Conclusions and outlook to future work are presented in \S~\ref{sec:conclusions}.

\section{Mathematical Formulation} \label{sec:formulation}

We focus on problems expressible in semi-discrete form as
\begin{equation}\label{eq:fom_ode}
  \frac{d \state}{dt}(t;\paramsA, \paramsF)
  =  \systemMat(\paramsA) \state(t;\paramsA, \paramsF) + \forcing(t;\paramsA,\paramsF),
\end{equation}
where $\state : [0,\tfinal] \times \paramDomainA \times \paramDomainF \rightarrow \RR{\fomDim}$
is the state, $\state(0;\paramsA, \paramsF) = \state^0$ is the initial state,
$\systemMat(\paramsA) \in \RR{\fomDim \times \fomDim}$ is the discrete system matrix
parametrized by ${\paramsA \in \paramDomainA \subset \RR{\nParamsA}}$,
$\forcing : [0,\tfinal] \times \paramDomainA \times \paramDomainF \rightarrow \RR{\fomDim}$
is the time-dependent forcing term also parametrized
by $\paramsF \in \paramDomainF \subset \RR{\nParamsF}$,
and $\tfinal>0$ is the final time.
We assume $\fomDim$ to be large, e.g., hundreds of thousands or more,
which is a suitable assumption for real applications.
Since the state is a rank-1 tensor (i.e., a vector), hereafter
we refer to a formulation of the form \eqref{eq:fom_ode} as
the rank-1 full-order model (\ronefom).

The above formulation is linear in the state, but can have an
arbitrary nonlinear parametric dependence.
No assumption is made on the original problem leading to \eqref{eq:fom_ode},
thus making it applicable to systems obtained from the spatial discretization
of a partial differential equation (PDE) or problems that are inherently discrete.
We assume the discrete system matrix, $\systemMat(\paramsA)$,
to be sparse, with its sparsity pattern depending on the problem and chosen discretization.
We remark that the formulation above, as well as what we present below,
is also applicable to the case where $\systemMat(\paramsA)$ has an affine parametrization.
We purposefully split the parametric dependence into $\paramsA$
and $\paramsF$ to highlight their different roles: $\paramsF$ includes
only the parameters that impact the forcing, while $\paramsA$
includes all the others.
This separation will be helpful for the formulations in the sections below.

\subsection{Applicability and representative applications}
A wide range of problems in science and engineering have governing
equations of the form \eqref{eq:fom_ode}. For example, the dynamics
of a deforming structure can often be modeled as linear, but the load
distribution on the structure can be parametrized by nonlinear
functions. This is typical in the field of aeroelasticity, where the
aircraft structure is modeled by a linear PDE and the aerodynamic loads
on the aircraft are highly nonlinear \cite{dowell2015aeroelasticity, meirovitch2010fundamentals, goman1994state, bazilevs2012isogeometric, borg2015frequency}.
Similarly, the temperature of a
structure may often be modeled with the linear heat equations and boundary
conditions set by nonlinear temperature distributions and/or heat loads \cite{incropera2007fundamentals}.
%% \EPC{The linear assumption is justified in cases with narrow operational
%% temperature ranges or materials with weak nonlinear effects.}
Nonlinear boundary conditions commonly arise in a number of applications
including electronics cooling \cite{zhao2002analysis, banerjee2009thermomagnetic}, building thermal management \cite{epting2017thermal}, and industrial
heat exchanger design \cite{shah2003fundamentals}.
Neutral particle (neutron, photon, etc.) transport, when simulated via
deterministic methods, often gives rise to linear systems of this form \cite{lewis1984computational, tencer2017accelerated}.
Acoustic waves are also modeled with a linear PDE, but can have a number
of nonlinear sources, most notably turbulent shear layers that arise in the
wakes of cars, aircraft, and other moving objects \cite{lighthill1952acoustic, fwh1969acoustic, phillips1960generation, lombard2015nonlinear}.
Linear circuit models are ubiquitous in electrical engineering and evolve
time according to ODEs of the form \eqref{eq:fom_ode} while commonly being
driven by complex nonlinear forcing functions \cite{vlach1983computer, chetty1982current, chen2005linear}.

\subsection{Galerkin ROM}
Projection-based ROMs generate approximate solutions of the full-order
model~\eqref{eq:fom_ode} by (a) restricting the state to live in a
low-dimensional (affine) \textit{trial} subspace, and (b) enforcing the residual of the
ODE to be orthogonal to a low-dimensional \textit{test} subspace.
In this work, we limit the scope to the Galerkin ROM,
which is one of the most popular methodologies (another one is, e.g., Petrov-Galerkin).

The Galerkin ROM approximates solutions
$\approxState(t;\paramsA, \paramsF) \approx \state(t;\paramsA, \paramsF)$
in a low-dimensional affine trial subspace,
${\approxState(t;\paramsA, \paramsF) \in \trialSubspace + \stateRef}$,
with ${\text{dim}(\trialSubspace) = K}$ such that $\romDim \ll \fomDim$
and where $\stateRef \in \RR{\fomDim}$
defines the affine offset (also called the reference state).
Here, we take $\trialSubspace = \text{Range}(\romBasis)$,
where ${\romBasis \equiv \begin{bmatrix} \romBasisCol_1 & \cdots & \romBasisCol_{\romDim } \end{bmatrix}}$
comprises $\romDim$-orthonormal basis vectors.
Such a basis may be obtained, for example, by proper orthogonal decomposition
\cite{sirovich1987,berkooz1993}.
For ${t \in [0,\tfinal]}$, ${\paramsA \in \paramDomainA}$, ${\paramsF \in \paramDomainF}$,
the FOM solution is approximated as
\begin{equation}\label{eq:pod_approx}
  \state(t;\paramsA, \paramsF) \approx \approxState(t;\paramsA, \paramsF)
  = \romBasis \romState(t;\paramsA, \paramsF) + \stateRef,
\end{equation}
where $\romState : [0,\tfinal] \times \paramDomainA \times \paramDomainF
\rightarrow \RR{\romDim}$ are referred to as the generalized coordinates.
The initial state for the generalized coordinates can be computed by
projecting the FOM initial condition, i.e., using
$\romState(0;\paramsA, \paramsF) = \romBasis^T \state(0;\paramsA, \paramsF)$.

\begin{figure}[!t]
  \centering
  \includegraphics[width=0.575\textwidth]{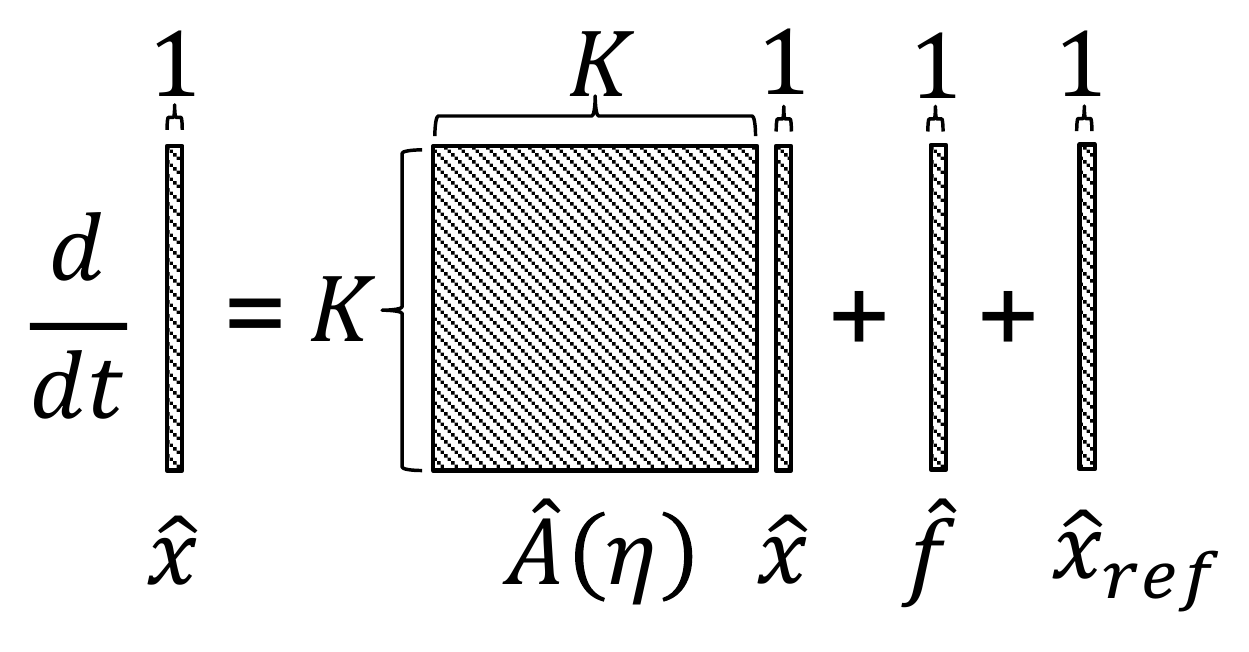}
  \caption{Schematic of the rank-1 Galerkin formulation discussed in Eq.~\ref{eq:rom_ode}.}
  \label{fig:formRank1}
\end{figure}

Equipped with~\eqref{eq:pod_approx}, the Galerkin ROM proceeds
by restricting the residual to be orthogonal to the trial space yielding
the following reduced-order model
\begin{equation}\label{eq:rom_ode}
  \frac{d \romState}{dt}(t;\paramsA, \paramsF) =
  \romSystemMat(\paramsA) \romState(t;\paramsA, \paramsF)
  + \hat{{\boldsymbol f}}(t;\paramsA, \paramsF)
  + \hat{{\boldsymbol x}}_\text{ref}(\paramsA, \paramsF)
\end{equation}
where $\romSystemMat(\paramsA) \equiv \romBasis^T \systemMat(\paramsA) \romBasis \in \RR{\romDim \times \romDim}$
is the reduced (dense) system matrix,
$\hat{{\boldsymbol f}}(t;\paramsA, \paramsF) = \romBasis^T \forcing(t;\paramsA, \paramsF)$
is the reduced forcing,
$\hat{{\boldsymbol x}}_\text{ref}(\paramsA, \paramsF) = \romBasis^T \systemMat(\paramsA) \stateRef$
is the reduced reference state, and we used the fact
that $\romBasis^T \romBasis = {\boldsymbol I}$.
A schematic of the above sysyem is shown in Figure~\ref{fig:formRank1}.
When the basis is not orthogonal, the transpose operations in
\eqref{eq:rom_ode} should be replaced with the pseudo-inverse.
Note that the affine offset, $\stateRef$, is introduced for generality,
but it not always necessary: when unused or null, it simply drops out of the formulation.
Hereafter we refer to a system of the form \eqref{eq:rom_ode} as
the rank-1 Galerkin ROM (\ronerom).

The last term on the right in \eqref{eq:rom_ode},
$\hat{{\boldsymbol x}}_\text{ref}(\paramsA, \paramsF) = \romBasis^T \systemMat(\paramsA) \stateRef$,
can be efficiently evaluated since it is time-independent and
only needs to be computed once for a given choice of $\paramsA$.
The term $\hat{{\boldsymbol f}}(t;\paramsA, \paramsF) = \romBasis^T \forcing(t;\paramsA, \paramsF)$,
despite being seemingly more challenging
because of its time-dependence, %and its complexity is $\ordOf{\fomDim}$.
can also be computed efficiently by considering two scenarios,
namely one where $\forcing \in \RR{\fomDim}$ is a dense vector
and one where it is sparse, with just a few non-zero elements.
In the former case, since $\romDim$ is sufficiently small,
the best approach would be to precompute and store
the product $\romBasis^T \forcing(t;\paramsA, \paramsF)$
for all the target times over $[0,\tfinal]$.
For the other scenario, i.e. a sparse forcing, one can exploit
its sparsity pattern at each time $t$ by operating only on
the corresponding rows of $\romBasis$. This allows one to avoid
pre-computing the forcing at all times while
maintaining computational efficiency since only a few elements must
be operated on.

%% The time integration of the reduced system is generally performed
%% using the same time stepping scheme, but not necessarily
%% the same time step size. In some cases, the reduced system can be integrated
%% in time using a larger time step size, which further improves the
%% computational efficiency, see \cite{Bach:2018}.
%% \EP{Why the same time stepping scheme? I wouldn't say that
%%   this is any more common or mathematically justified than different time step sizes}

\subsection{Rank-1 Formulation Analysis}
The \ronefom{} and \ronerom{} formulations are suitable when the objective
is to perform a few individual cases, but become inefficient for many-query
scenarios involving large ensembles of runs on modern architectures.
The reason is that computing the right-hand side of both
\ronefom{} and \ronerom{} requires memory bandwidth-bound kernels,
which impede an optimal/full utilization of a node's computing resources,
especially on modern multi- and many-core
computer architectures \cite{Hutcheson2011MemoryBV, yuen2011, elafrou2017}.
A partial solution would be to run the simulations in parallel,
as demonstrated in \cite{yang2017grom}, but the individual runs
in this approach would still be of the same nature.

Indeed, the \ronefom{} in \eqref{eq:fom_ode} is characterized by
a standard \underline{sp}arse-\underline{m}atrix \underline{v}ector
(\code{spmv}) product, which is well-known to be memory bandwidth bound
due to its low compute intensity regardless
of its sparsity pattern, see e.g. \cite{Bell:SPMV:2008, elafrou2017}.
Defining the computational intensity, $I$, of a kernel
as the ratio between flops and memory access (bytes),
the \code{spmv} kernel has approximately $I \approx nnz/(6\fomDim + 2 + 10nnz)$,
where $nnz$ is the number of non-zero elements \cite{Bell:SPMV:2008, elafrou2017}.
%Achieving high performance for sparse kernels is non-trivial
%due to the irregular data access patterns and is an active thrust of research.

For the \ronerom{} in \eqref{eq:rom_ode}, the defining term is
the product $\romSystemMat(\paramsA) \romState(t;\paramsA, \paramsF)$
of the reduced system matrix with the vector of generalized coordinates,
which requires a dense \underline{ge}neral \underline{m}atrix
\underline{v}ector (\code{gemv}) product.
This kernel is also memory bandwidth bound with an approximate
computational intensity $I \approx 1/4$ \cite{peise2017},
when the matrix is square as in \eqref{eq:rom_ode}.

%% one are typically solved multiple times
%% by analysts to compute a set of $\nRuns$ trajectories
%% $\approxState_i(t;{\paramsA}_i, {\paramsF}_i)$ for
%% ${\paramsA}_i, {\paramsF}_i \in \paramDomainA \times \paramDomainF$, $i=1,\ldots,\nRuns$.

\subsection{Rank-2 Formulation}
In light of the previous discussion, we now present an alternative
formulation and implementation that is computationally more efficient
for many-query problems with respect to the parameter $\paramsF$ on modern many-core architectures.
Let $\stateTensor : [0,\tfinal] \times \paramDomainA \times \paramDomainF^{\nRuns}
\rightarrow \RR{\fomDim \times \nRuns}$
represent a set of $\nRuns$ trajectories for the FOM
\begin{equation}
\stateTensor(t;\paramsA, \paramsMat) \equiv
\begin{bmatrix}
\state_1(t;\paramsA, {\paramsF}_1) & \hdots & \state_{\nRuns}(t;\paramsA, {\paramsF}_{\nRuns})
\end{bmatrix},
\notag
\end{equation}
for a given choice of $\paramsA$ and where
$\paramsMat = {\paramsF}_1, \ldots, {\paramsF}_{\nRuns}$
is the set of parameters defining the $\nRuns$ forcing realizations
driving the trajectories of interest.

Similarly to~\eqref{eq:fom_ode}, we can express the dynamics
of these FOM trajectories as
\begin{equation}\label{eq:fom_ode_tensor}
  \frac{d \stateTensor}{dt}(t;\paramsA, \paramsMat)
  = \systemMat(\paramsA) \stateTensor(t;\paramsA, \paramsMat)
  + \forcingTensor(t;\paramsA, \paramsMat),
\end{equation}
where
$\stateTensor(0;\paramsA, \paramsMat) =
\begin{bmatrix}
  \state_1(0;\paramsA, {\paramsF}_1)
  & \cdots
  & \state_{\nRuns}(0;\paramsA, {\paramsF}_{\nRuns})
\end{bmatrix}$
is the initial condition, and the forcing contribution is
$\forcingTensor (t;\paramsA, \paramsMat) \equiv
\begin{bmatrix}
  \forcing(t;\paramsA, {\paramsF}_1) & \cdots & \forcing(t; \paramsA,{\paramsF}_{\nRuns})
\end{bmatrix}$.
We refer to the system of the form \eqref{eq:fom_ode_tensor}
as the rank-2 FOM (\rtwofom). The \ronefom{} in \eqref{eq:fom_ode} can be seen
as a special case of Eq.~\eqref{eq:fom_ode_tensor} with $\nRuns=1$.
%In the above $\forcingTensor : [0,\tfinal] \times \paramDomainF^{\nParamsF}$

\begin{figure}[!t]
  \centering
  \includegraphics[width=0.8\textwidth]{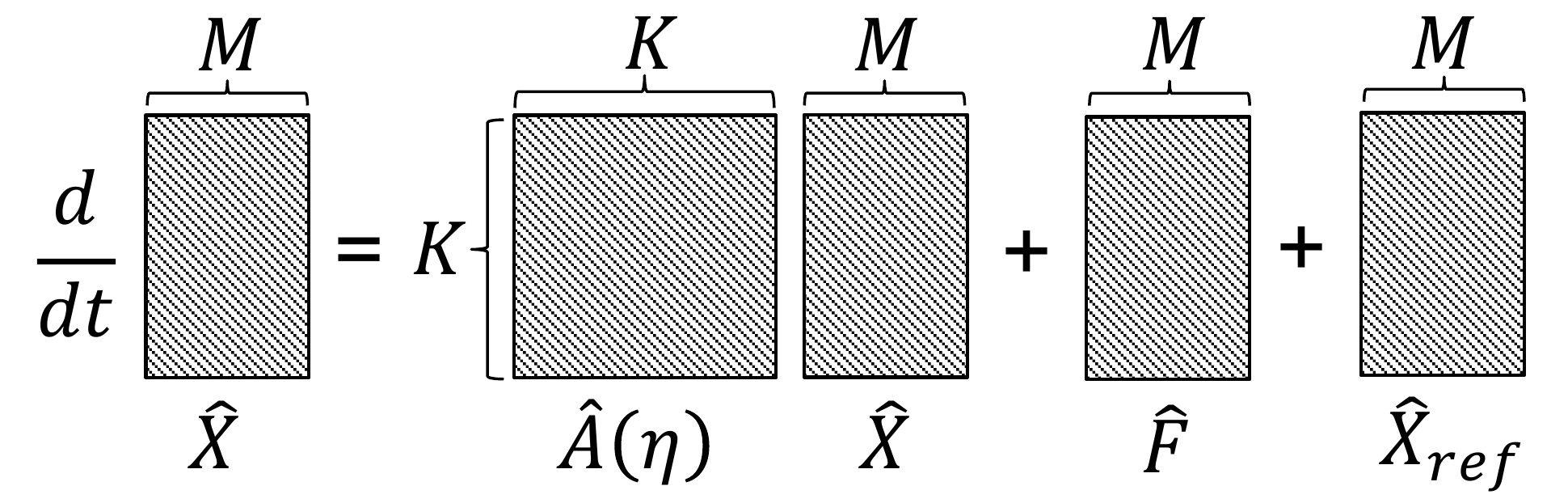}
  \caption{Schematic of the rank-2 Galerkin formulation discussed in Eq.~\ref{eq:rom_ode_tensor}.}
  \label{fig:formRank2}
\end{figure}

To derive the corresponding Galerkin ROM, we approximate the state as
\begin{equation}
  \stateTensor(t;\paramsA, \paramsMat)
  \approx \approxStateTensor(t;\paramsA, \paramsMat) \equiv
  \romBasis \romStateTensor(t;\paramsA, \paramsMat)
  + \stateTensorRefWithPars, \notag
\end{equation}
where
$\stateTensorRefWithPars \equiv \begin{bmatrix}
  \stateRefNoPars(\paramsA, {\paramsF}_1 ) & \cdots & \stateRefNoPars(  \paramsA, {\paramsF}_{\nRuns}) \end{bmatrix}
\in \RR{N \times M}$,
and applying Galerkin projection we obtain
\begin{equation} \label{eq:rom_ode_tensor}
  \frac{d \romStateTensor}{dt}(t;\paramsA, \paramsMat)
  = \romSystemMat(\paramsA) \romStateTensor(t;\paramsA, \paramsMat)
  + \hat{{\boldsymbol \forcingTensor}}(t;\paramsA, \paramsF)
  + \hat{{\boldsymbol X}}_\text{ref}(\paramsA, \paramsMat),
\end{equation}
where
$\romStateTensor : [0,\tfinal] \times \paramDomainA \times
\paramDomainF^{\nRuns} \rightarrow \RR{\romDim \times \nRuns}$
is a rank-2 tensor of time-dependent generalized coordinates,
$\hat{{\boldsymbol \forcingTensor}}(t;\paramsA, \paramsF) = \romBasis^T \forcingTensor(t;\paramsA, \paramsMat)$
is the reduced forcing,
$\hat{{\boldsymbol X}}_\text{ref}(\paramsA, \paramsMat) = \romBasis^T \systemMat(\paramsA) \stateTensorRefWithPars$
is the reduced reference state.
Figure~\ref{fig:formRank2} shows a schematic of the system in \eqref{eq:rom_ode_tensor},
which hereafter we refer to the rank-2 Galerkin ROM (\rtworom).
The term $\romBasis^T \systemMat(\paramsA) \stateTensorRefWithPars$
can be efficiently evaluated since it is time-independent and
only needs to be computed once for a given choice of $\paramsA$.
For the term $\romBasis^T\forcingTensor(t;\paramsA, \paramsMat)$,
involving the forcing, considerations similar to those drawn for
the rank-1 Galerkin in Eq.~\eqref{eq:rom_ode} can be made.
The term involving the system matrix is discussed below in more detail.

\subsection{Rank-2 Formulation Analysis}
The key question we now pose is what advantage, if any, is gained by
employing the \mbox{rank-2} formulation of the FOM and Galerkin ROM
over the \mbox{rank-1} alternatives?
The answer is that, when evaluated in a many-query context,
\rtwofom{} has minimal advantages over \ronefom,
whereas \rtworom{} has major benefits over \ronerom.
We will demonstrate this numerically in \S~\ref{sec:fomScaling}
and \S~\ref{sec:romScaling}, but provide the key insight here.

For the FOM, the \rtwofom{} in \eqref{eq:fom_ode_tensor}
involves the term
$\systemMat(\paramsA) \stateTensor(t;\paramsA, \paramsMat)$
requiring a \underline{sp}arse-\underline{m}atrix \underline{m}atrix
(\code{spmm}) kernel which, despite having a compute intensity
higher than \code{spmv}, remains memory bandwidth bound,
see e.g. \cite{khalid:2017, hong:2018}.
This implies that \rtwofom{} yields some but limited
improvements over \ronefom.

On the other hand, the \rtworom{} in \eqref{eq:rom_ode_tensor}
has a major advantage over \ronerom{} because it changes the
nature of the problem from memory bandwidth to compute bound.
This stems from the fact that computing the term
$\romSystemMat(\paramsA) \romStateTensor(t;\paramsA, \paramsMat)$
on the right-hand side of \rtworom{} in \eqref{eq:rom_ode_tensor}
involves a dense \underline{ge}neral \underline{m}atrix \underline{m}atrix
multiplication (\code{gemm})\footnote{The \code{gemm} kernel
is ${\boldsymbol C} = \alpha \cdot op({\boldsymbol A}) op({\boldsymbol B})
+ \beta \cdot {\boldsymbol C}$ where $\alpha, \beta$ are scalars, ${\boldsymbol A}, {\boldsymbol B}$
and ${\boldsymbol C}$ are matrices, and $op({\boldsymbol X})= {\boldsymbol X}$
or $op({\boldsymbol X})={\boldsymbol X}^T$.},
which has the advantage of
performing $\mathcal{O}(n^3)$ flops on $\mathcal{O}(n^2)$ memory accesses,
yielding a compute bound kernel (\cite{peise2017}).
The \code{gemm} is one the most studied kernels in dense linear algebra.
The approximate computational intensity for \code{gemm}
is $I \approx \romDim/16$, since the reduced system matrix
in Eq.~\eqref{eq:rom_ode_tensor} is square with size $\romDim \times \romDim$.
%% The \code{gemm} kernel is at the core of level-3 BLAS,
%% and over the years has been the focus of considerable work to optimize it.
%% Consequently, by relying on it, one can benefit from
%% all the work completed over the years towards optimizing it.
This makes the \mbox{rank-2} Galerkin ROM very well-suited for
modern multi- and many-core computing nodes where a high computational
intensity is critical for achieving good scaling and efficiency.

Previous work in the field of computational science that
is relevant can be found, e.g., in \cite{Jiang_2014, gunz2017, gunz2018, phipps2015embedded}.
In \cite{phipps2015embedded}, the authors propose an ``embedded ensemble propagation''
technique which exploits modern CPUs' vectorization and C++ templates to
simultaneously propagate multiple samples together. This work, however, does not explicitly address ROMs.
In \cite{Jiang_2014, gunz2017}, the authors focus on
efficient reformulations of the time-dependent Navier-Stokes equations to solve ensenbles.
The idea is to employ a reformulation that allows to solve a single linear system for multiple right-hand sides.
These studies, however, are tailored to the Navier-Stokes equations.
The work in \cite{gunz2017} is based on the previous ones, but is particularly
relevant to model reduction because the authors discuss an ensemble-based POD Galerkin
ROM tailored to the incompressible Navier—Stokes equations for efficiently solving
problems characterized by multiple initial conditions and body forces.
%%The authors also discuss how to extend this to  idea is to exploit the quadratic nonlinearity
%% in the convection term and define an ensemble mean on the velocity that
%% enables one to solve the Navier—Stokes equations for
%% all ensemble members by solving a multiple RHS linear system at each time step.
The idea is based on first performing an ensemble reformulation of the FOM,
from which a POD-Galerkin ROM is then constructed. The resulting method is first-order accurate in time, and is tailored to the Navier-Stokes equations.
We note that no performance results or comparisons to standard approaches are presented in \cite{gunz2017}.

\section{Test Case Description} \label{sec:testcase}

As a test case for our numerical experiments, we choose the propagation
of global seismic shear waves in spherical coordinates.
Shear waves are also called S-waves (or secondary)
because they come after P-waves (or primary).
The main difference between them is that S waves
are transversal (particles oscillate perpendicularly
to the direction of wave propagation), while the P waves
are longitudinal (particles oscillate in the same direction as the wave).
Both P and S waves are body waves, because they travel through
the interior of the earth and their trajectories are affected
by the material properties, namely density and stiffness.

We specifically focus on the axisymmetric
evolution of elastic seismic shear waves.
The governing equations in the velocity-stress
formulation \cite{igel1995, shaxi2008} can be written as:
\begin{align}\label{eq:fom_asym}
  \dens (\coords) \dvpdt (\coords,t) &=
  \dsrpdr(\coords,t) + \frac{1}{r} \dstpdtheta(\coords,t)\notag\\
  &+ \frac{1}{r} \left(3 \srp(\coords,t) + 2 \stp(\coords,t) \cot{\theta} \right)
  + \force(\coords, t) \notag\\
  \dsrpdt(\coords,t) & = \shMod(\coords) \left( \dvpdr(\coords,t) - \frac{1}{r}\vp(\coords,t) \right)\\
  \dstpdt(\coords,t) & = \shMod(\coords) \left( \frac{1}{r}\dvpdtheta(\coords,t)
  - \frac{\cot{\theta}}{r}\vp(\coords,t) \right), \notag
\end{align}
where $t$ represents time, $r \in [0, r_{earth}]$ is the radial distance bounded
by the radius of the earth, $\theta \in [0, \pi]$ is the polar angle,
$\dens(\coords)$ is the density,
$\vp(\coords, t)$ is the velocity, $\srp(\coords, t)$ and $\stp(\coords, t)$ are
the two components of the stress tensor remaining after the
axisymmetric approximation, $f(\coords,t)$ is the forcing term,
and the shear modulus is $\shMod(\coords) = v_s^2(\coords) \dens(\coords)$,
with $v_s$ being the shear wave velocity.
Note that we assume both the density and shear modulus to only depend
on the spatial coordinates.
Such a formulation is referred to as \mbox{2.5-dimensional} because
it involves a 2-dimensional spatial domain (a circular sector of the earth)
but models point sources with correct 3-dimensional spreading \cite{igel1995, shaxi2008}.

This is an application of interest to geological
and, in particular, seismic modeling. Previous studies focusing
on shear waves modeling can be found, e.g.,
in \cite{igel1995, chaljub1997, shaxi2008, wong2012} and
references therein.
%In particular, we acknowledge the work by Jahnke {\it et.~al.}~\cite{shaxi2008},
%since the authors derive a high-order
%finite-difference scheme for simulating elastic shear waves in axisymmetric
%media for parallel computers with distributed memory, and
%use it primarily to study the effects of perturbations
%in the material model.

\subsection{Discretization}
\begin{figure}[!t]
  \centering
  %% \begin{minipage}{0.5\textwidth}
  %% \input{spher_coords}
  %% \end{minipage}
  %\begin{minipage}{0.48\textwidth}
  \includegraphics[width=0.4\textwidth]{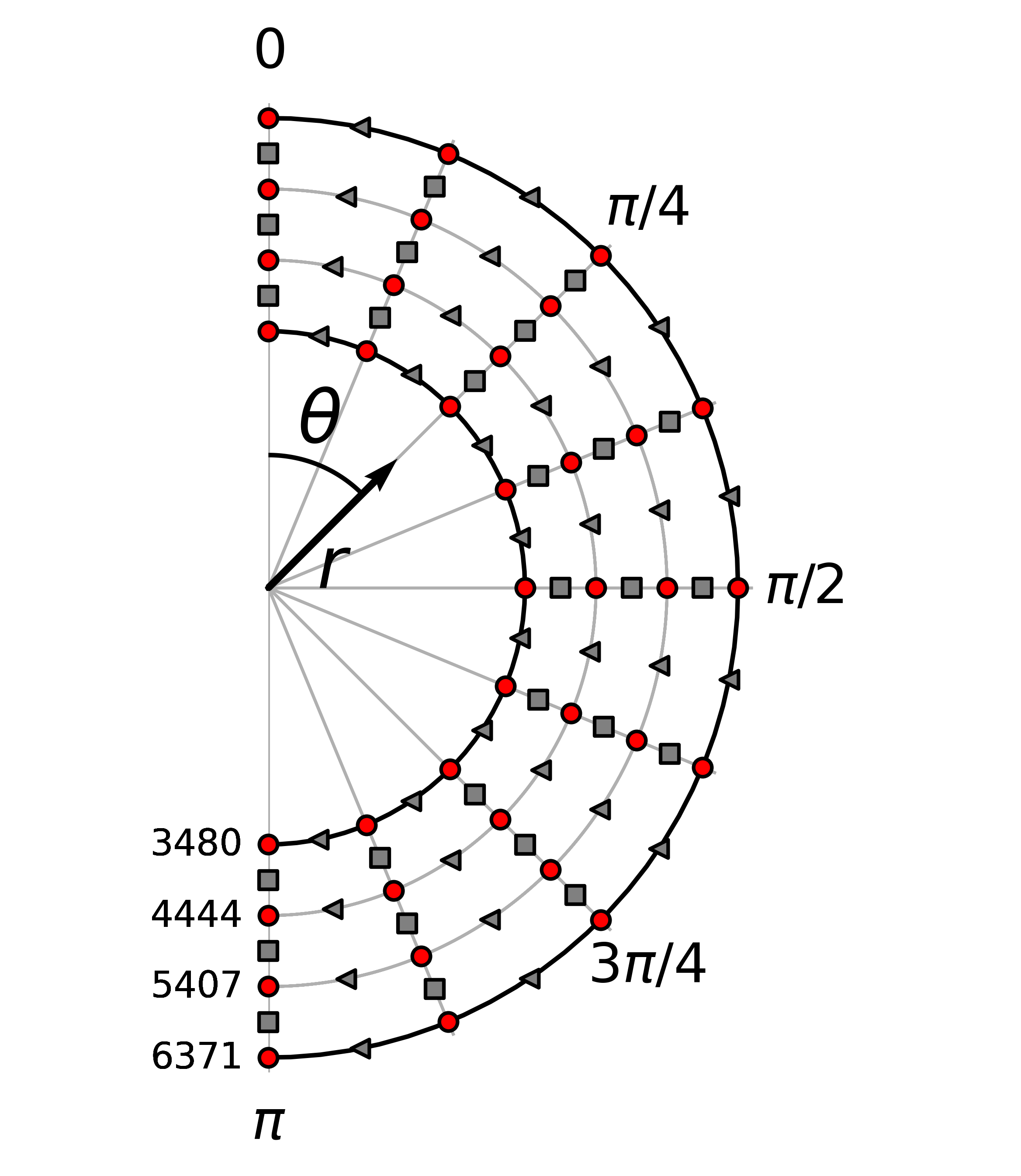}(a)
  \includegraphics[width=0.4\textwidth]{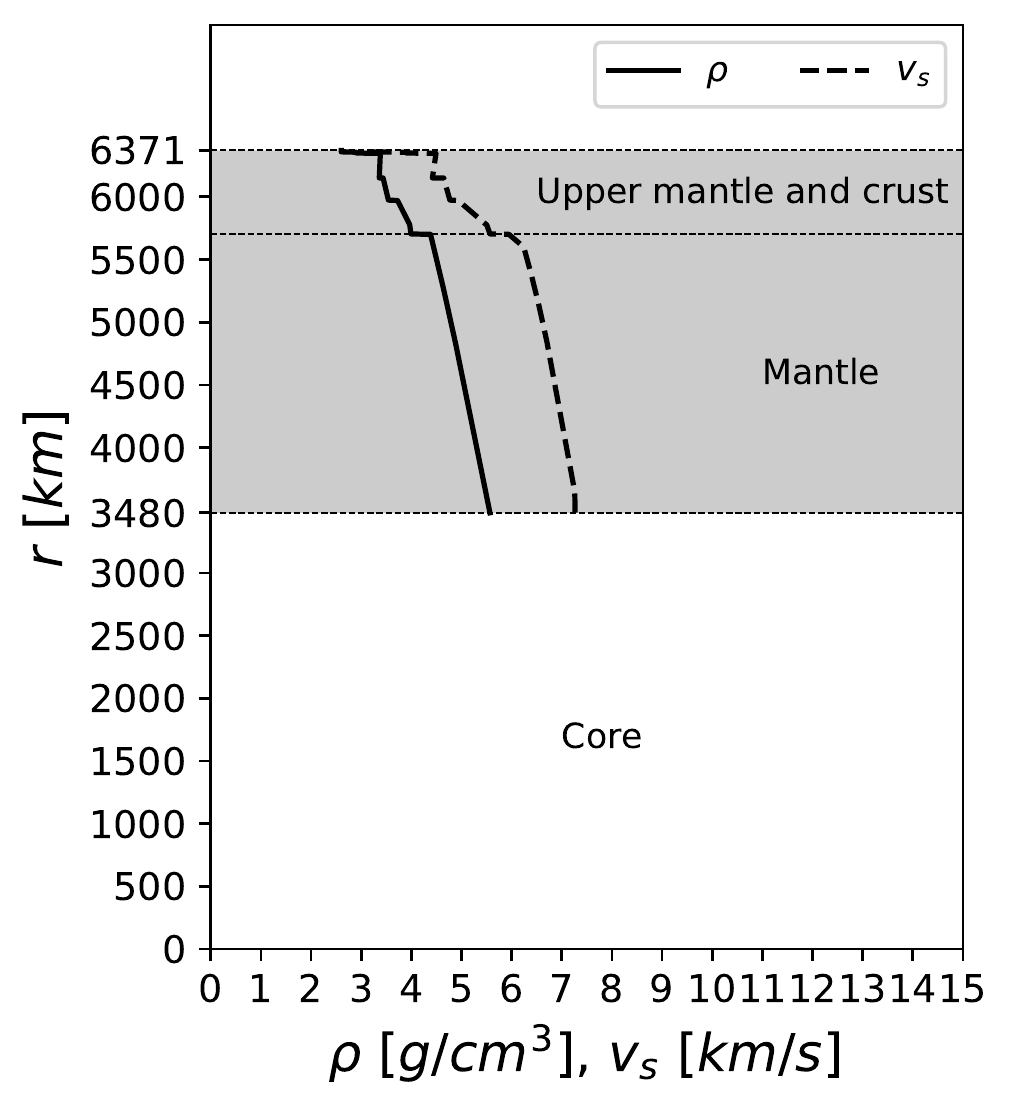}(b)
  %\end{minipage}
  \caption{(a): schematic of the axi-symmetric domain and
    staggered grid used for simulating the shear wave
    system in Eq.\eqref{eq:fom_asym},
    where the red-filled circles represent grid points
    where the velocity $\vp$ is defined,
    the gray-filled squares represent those
    for the stress $\srp$ and the gray-filled triangles are
    those for the stress $\stp$.
    (b): profiles of density $\rho$ and shear velocity $v_s$
    as a function of the radial distance down to the core-mantle boundary
    as defined by the PREM model.
    The shaded gray region represents the simulation domain.}
  \label{fig:mesh}
\end{figure}

Since shear waves cannot propagate in liquids, the system of equations
in \eqref{eq:fom_asym} is not applicable to the core region of the earth.
Therefore, we solve Eq.~\eqref{eq:fom_asym} in the region bounded between the core-mantle
boundary (CMB) located at $r_{cmb}=3,480$~km, and the earth surface located
at $r_{earth}=6,371$~km. We discretize the 2D spatial domain using
a staggered grid as shown in Figure~\ref{fig:mesh}~(a).
Staggered grids are typical for seismic modeling
and wave problems in general, see \cite{shaxi2008, virieux1984, virieux1986}.
We use a second-order centered finite-difference \cite{shaxi2008} method 
for the spatial operators in Eq.~\eqref{eq:fom_asym}, which is sufficient 
for the purposes of this study.

The \rtwofom~formulation of Eq.\eqref{eq:fom_asym} can be written as
\begin{align}
  \label{eq:shwave_ode_tensor}
  \frac{d \stateTensor_{\vp}}{dt}(t;\paramsA, \paramsMat)
  &= \systemMat_{\vp}(\paramsA) \stateTensor_{\stresses}(t;\paramsA, \paramsMat)
  + \forcingTensor_{\vp}(t;\paramsA, \paramsMat) \notag\\
  \frac{d \stateTensor_{\stresses}}{dt}(t;\paramsA, \paramsMat)
  &= \systemMat_{\stresses}(\paramsA) \stateTensor_{\vp}(t;\paramsA, \paramsMat),
\end{align}
where $\stateTensor_{\vp} \in \RR{\fomDim_{\vp} \times \nRuns}$ is the
rank-2 state tensor for the velocity degrees of freedom,
$\stateTensor_{\stresses} \in \RR{\fomDim_{\stresses} \times \nRuns}$ is the
rank-2 state tensor with the stresses degrees of freedom,
$\systemMat_{\vp}(\paramsA) \in \RR{\fomDim_{\vp} \times \fomDim_{\stresses}}$
is the discrete system matrix for the velocity,
$\systemMat_{\stresses}(\paramsA) \in \RR{\fomDim_{\stresses} \times \fomDim_{\vp}}$
is the one for the stresses, and $\forcingTensor_{\vp}(t;\paramsA, \paramsMat)$ is the
rank-2 forcing tensor.
In this work, the finite difference scheme adopted leads to
system matrices $\systemMat_{\vp}(\paramsA)$ and $\systemMat_{\stresses}(\paramsA)$
with about four and two non-zeros entries per row, respectively.
The corresponding \ronefom~can be easily obtained
from \eqref{eq:shwave_ode_tensor} by setting $\nRuns=1$.

%We complement the governing equations with the following boundary conditions.
We use the following boundary conditions, similarly to \cite{shaxi2008}.
Directly at the symmetry axis, i.e., $\theta=0$ and $\theta=\pi$,
the velocity is set to zero since it undefined here due
to the cotangent term in its governing equation. This implies that
at the symmetry axis the stress $\srp$ is also zero.
At the core-mantle boundary and earth surface we impose a free
surface boundary condition (i.e., waves fully reflect),
by setting the zero-stress condition 
${\srp(r_{cmb}, \theta, t)=\srp(r_{earth}, \theta, t)=0}$.
Note that this condition on the stress implicitly
defines the velocity at the core-mantle boundary
and earth surface and, therefore, no boundary condition
on the velocity itself must be set there.
We remark that, differently than \cite{shaxi2008},
we do not rely on ghost points to impose boundary conditions,
because these are already accounted
for when assembling the system matrices $\systemMat_{\vp}(\paramsA)$ and
$\systemMat_{\stresses}(\paramsA)$.

%% Hence, we obtain a system of two coupled equations
%% of the form Eq.~\eqref{eq:fom_ode} for rank-1 FOM and of the
%% form Eq.~\eqref{eq:fom_ode_tensor} for the rank-2 FOM, one
%% for the velocity and one for the stresses.
The \rtworom~formulation of the system in Eq.\eqref{eq:shwave_ode_tensor}
can be expressed as
\begin{align} \label{eq:shwave_rom_ode_tensor}
  \frac{d \romStateTensor_{\vp}}{dt}(t;\paramsA, \paramsMat)
  &= \romSystemMat_{\vp}(\paramsA) \romStateTensor_{\stresses}(t;\paramsA, \paramsMat)
  + \romBasis_{\vp}^T\forcingTensor_{\vp}(t;\paramsA, \paramsMat)
  %+ \romBasis_^T \systemMat(\paramsA) \stateTensorRef
  \notag\\
  \frac{d \romStateTensor_{\stresses}}{dt}(t;\paramsA, \paramsMat)
  &= \romSystemMat_{\stresses}(\paramsA) \romStateTensor_{\vp}(t;\paramsA, \paramsMat),
\end{align}
where
$\romSystemMat_{\vp}(\paramsA) \equiv
\romBasis_{\vp}^T \systemMat_{\vp}(\paramsA) \romBasis_{\stresses}
\in \RR{\romDim_{\vp} \times \romDim_{\stresses}}$
is the reduced system matrix for the velocity and
$\romSystemMat_{\stresses}(\paramsA) \equiv
\romBasis_{\stresses}^T \systemMat_{\stresses}(\paramsA) \romBasis_{\vp}
\in \RR{\romDim_{\stresses} \times \romDim_{\vp}}$
is the one for the stresses, $\romBasis_{\vp}$ and $\romBasis_{\stresses}$
are the orthonormal (i.e., $\romBasis_{\vp}^T \romBasis_{\vp} = \mathbf{I}$
and the same for $\romBasis_{\stresses}$)
bases for the velocity and stresses, respectively.
This enables, by default, a physics-based separation of the degrees
of freedom which inherently have considerably disparate scales.
This benefits the computation of the POD modes for the ROM,
because they can be computed independently for velocity and stresses,
and no specific scaling is required. Furthermore, it allows one
to potentially use different basis sizes to represent
the velocity and stresses degrees of freedom.
Note that we omitted the reference state because we do not use it here.
The corresponding \ronerom~can be obtained by setting $\nRuns=1$.

For time integration of both the FOM and ROM, we use a leapfrog integrator,
where the velocity field is updated first, followed by the stress update.
This is a commonly used scheme for classical mechanics because it is
time-reversible and symplectic. The initial conditions consist
of zero velocity and stresses.

To the best of our knowledge, this work provides the first
example of ROMs applied to elastic shear wave simulations.
However, it is important to acknowledge that the application
of ROMs to wave propagation is not new, and refer the reader
to \cite{pereyra2008, pereyra2016}.
These two studies focused on illustrating the feasibility and
quantifying the accuracy of ROMs for acoustic wave simulations.
Of particular interest is the discussion presented in \cite{pereyra2016}
on various techniques to make ROMs feasible for three-dimensional
acoustic simulations, a non-trivial task due to the size of the problem.
In a future study, one could explore, for example, the combination
of some of the methods presented in \cite{pereyra2008} with
our rank-2 formulation.

\subsection{Implementation, Material Model and Forcing}

We implemented the rank-1 and rank-2 formulations
of both the FOM and ROM for the shear problem above 
in C++ using the performance-portable 
Kokkos programming model~\cite{CarterEdwards:2014}\footnote{\url{https://github.com/kokkos/kokkos}}, 
and make the repository publicly available 
at \url{github.com/fnrizzi/ElasticShearWaves} to favor 
scientific reproducibility and benefit other researchers.
Kokkos exposes a general API allowing users to write code only once,
and internally exploits compile-time polymorphism to
decide on the proper data layout, and dispatching to the
suitable backend depending on the target threaded system,
e.g.,~OpenMP, CUDA, OpenCL, etc.
Therefore, once the code is written, it can be easily tested
on CPUs and their corresponding GPUs with minimal effort.

One of the key aspects related to modeling seismic shear waves
is the choice
of material model (i.e., density and shear modulus) and forcing term.
Our code currently supports a few 1D material models:
a single layer and a bilayer model and the Preliminary Reference
Earth Model (PREM)~\cite{prem:1981}.
We are in the process of adding support for the
ak135f~\cite{kennett:1995, montagner:1996}\footnote{\url{http://rses.anu.edu.au/seismology/ak135/ak135f.html}}.
We remark that these are 1D models in the sense that they only
depend on the radial distance.
For the single layer model, one needs to set
the radial profile of the density and shear wave velocity.
For the bilayer model, one needs to set the depth of
the discontinuity separating the layers, and profiles within each layer.
The single and bilayer models are useful for testing and
exploratory cases but are obviously not fully representative of the earth structure.
The PREM and ak135f are two of the most commonly used models.
These models have several discontinuities and complex profiles
describing the
material properties \footnote{\url{https://seismo.berkeley.edu/wiki\_cider/REFERENCE\_MODELS}}.
Figure~\ref{fig:mesh}~(b) shows the profiles of the material
properties as defined by the PREM model.
We designed the code in a modular fashion such that adding new
material models or extending existing ones is straightforward.
%We anticipate that the results presented in this
%work are based on using the PREM model.
%% , depending on which order is used for
%% the polynomial expressing the profile of density
%% and shear velocity within each layer of the model.
%% Each material model is indeed
%% implementend as a stand alone functor that does only needs to
%% expose the parens operator where, given a radius and a theta,
%% it computes the corresponding density and shear velocity.
%% The shear modulus is then easily computed as in equation ...
%% This modularity is useful for several reasons: one can explore different
%% variants as well possibly exploring model of the moon or other
%% planets by using data without needing to touch or know
%% the implementation details of the rest of the code.

%If the objective is to simulate the effect of shear waves generated by
%earthquakes, the source term should be a realistic model of an earthquake excitation.
The forcing term has two main defining aspects,
the form of the excitation signal and the location at which it acts.
We support two kinds of signals commonly used for seismic modeling,
namely a Ricker wavelet and the first derivative of
a Gaussian~\cite{Rabinovich:2018}.
The main parameter to set for these signals is the period, $T$,
which is a key property to explore for characterizing seismic events.
Similar to the design of the material model, our implementation is easily
extended to include other kinds of signals.
As for the location, due to the axisymmetric approximation,
the seismic source cannot be placed exactly on the
symmetry axis \cite{shaxi2008}, i.e. $\theta=0$ and $\theta=\pi$.
Therefore, for a given depth, it is placed at
the first velocity grid point such that $\theta>0$.

\section{Results} \label{sec:results}
The results presented below are obtained with a machine
containing two 18-core Intel(R) Xeon(R) Gold 6154 CPU @ 3.00GHz, each with
a 24.75MB L3 cache and 125GB total memory. 
We enable hyperthreading, thus supporting a maximum 
of 36 logical threads per CPU, so a total of 72 threads.
We use GCC-8.3.1 and rely on kokkos and kokkos-kernels version 3.1.01.
We use Blis-0.7.0~\cite{BLIS1, BLIS2} as the kokkos-kernels' backend
for all dense operations.
In the present work we only run on CPUs enabling OpenMP as
the default Kokkos execution space.
A more detailed study encompassing the role of the compiler and architecture
(e.g., running on GPUs) is left for a future work.
The results are divided into four parts. Scaling and performance results
obtained for the FOM are presented in \S~\ref{sec:fomScaling} and
for the Galerkin ROM in \S~\ref{sec:romScaling}.
Accuracy results and a representative use case for a many-query problem
are discussed for the ROM in \S~\ref{sec:romAccuracy}
and \S~\ref{sec:mqRom}, respectively.
For reproducibility purposes, in the supplementary material we provide
the full details to access the data corresponding to the results
below as well as the scripts used for the runs.

\subsection{Full-order Model Performance and Scaling}
\label{sec:fomScaling}

In this section, we present scaling and performance tests obtained
for the shear wave simulations using the \ronefom~and \rtwofom,
see Eq.\eqref{eq:shwave_ode_tensor}.
For brevity, we omit the full details of the model and physical
parameters used to carry out these FOM numerical experiments and refer
the reader to supplemental material. For this analysis
the physical details are not important because they
only come into play during the preprocessing stage
and, therefore, have no impact on the performance.

%We are interested in exploring the impact of the problem size, number of threads and forcing size.
We consider $\nRuns \in \{1,2,4,8,16,32,48\}$,
thread counts $2,4,8,12,18,36,72$, and total degrees
of freedom $\fomDim \in \{785,152; 3,143,168; 12,577,792; 50,321,408\}$.
These values of $\fomDim$ are the total degrees of freedom
originating from choosing the following grids for the velocity:
$256 \times 1024$, $512 \times 2048$, $1024 \times 4096$ and $2048 \times 8192$.
%Note that for the configurations chosen, cache effects are mainly
%affecting only the first two values of $\fomDim$.
Note that in general one should choose the value of $\nRuns$
considering its trade off with the amount of memory required.
Indeed, if one needs to save state data very frequently, using
large values of $\nRuns$ can yield a very large memory utilization
for the FOM problem.
Here, to make this FOM analysis feasible for $\nRuns \leq 48$
and since we are not interested in the physical data,
we disable all input/output related to saving data.

For the \rtwofom, we use a row-major ordering for the state matrix.
This choice, given the OpenMP execution space chosen for Kokkos,
yielded a performance superior to using column-major ordering.
%% In this study, we use row-major ordering (also called layout-right in Kokkos).
%% For the \code{spmm} kernel involved in the \rtwofom, combining
%% a row-major layout for the state with the OpenMP backend of Kokkos-kernels
%Note, however, that if we were to run the code on CUDA,
%the same setup might not be as performant, and the column-major
%layout might be a better choice.
For the sake of clarity, we remark that for the \ronefom{} we
do not need to choose the memory layout, because a rank-1 state is just a 1D array.
For both \ronefom~and \rtwofom{} we use the following OpenMP
affinity\footnote{\url{https://www.openmp.org/spec-html/5.0/openmp.html},
\url{https://github.com/kokkos/kokkos-kernels/wiki/spmv_becnhmarks}}:
\code{OMP\_PLACES=threads} and \code{OMP\_PROC\_BIND=spread}.
\begin{figure}[!t]
  \centering
  \includegraphics[width=0.65\textwidth]{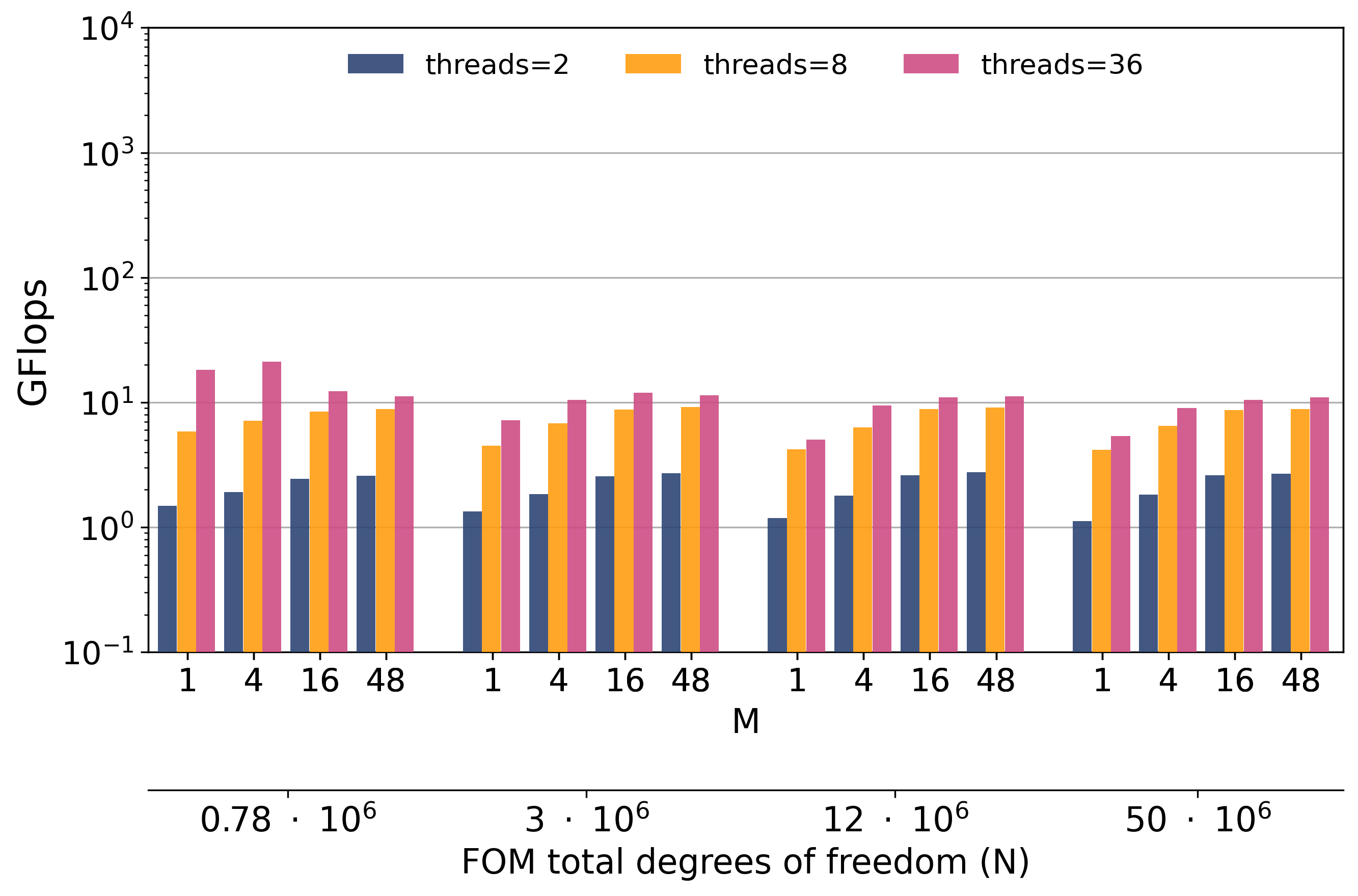}(a)\\
  \includegraphics[width=0.65\textwidth]{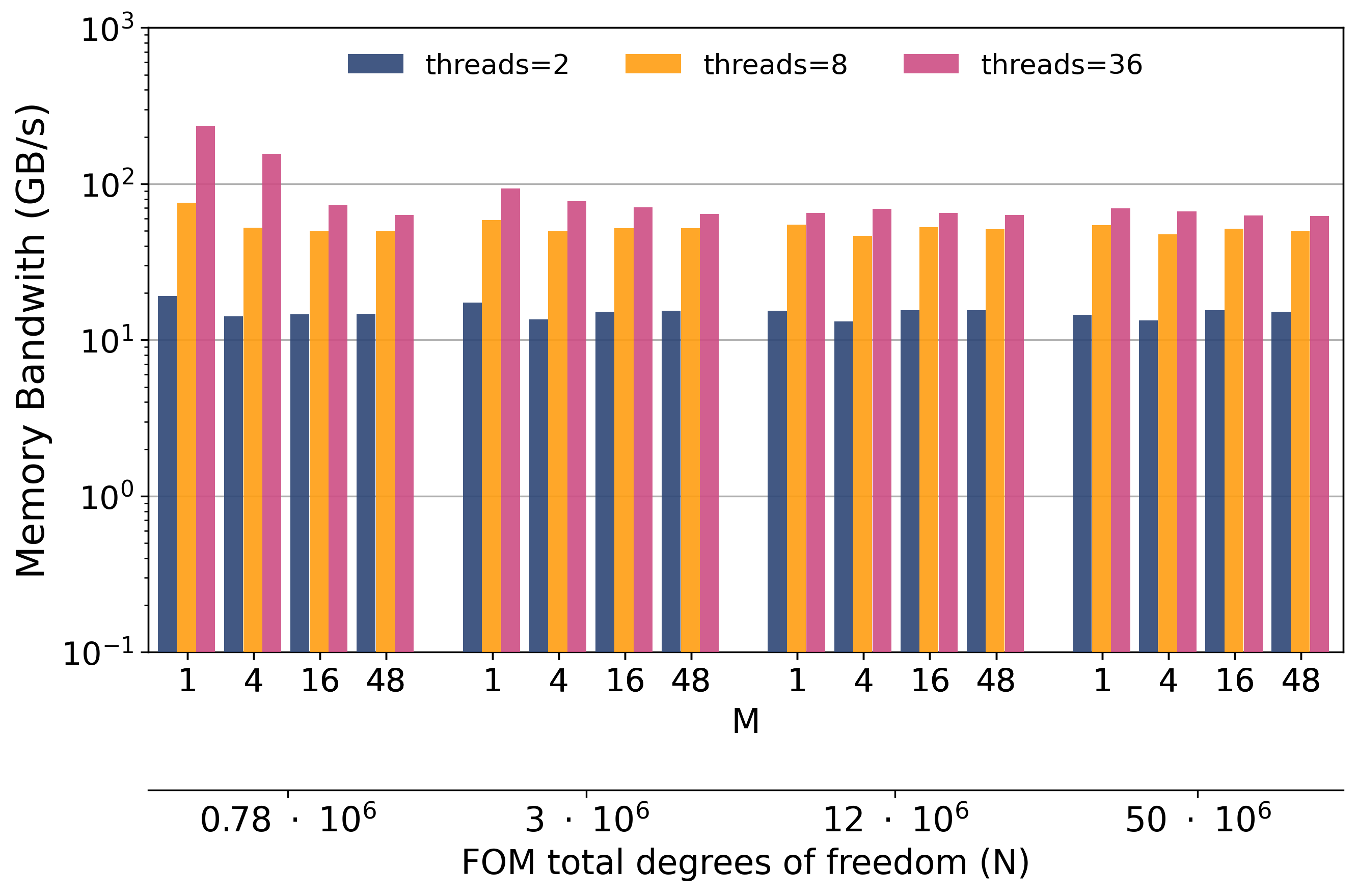}(b)\\
  \includegraphics[width=0.65\textwidth]{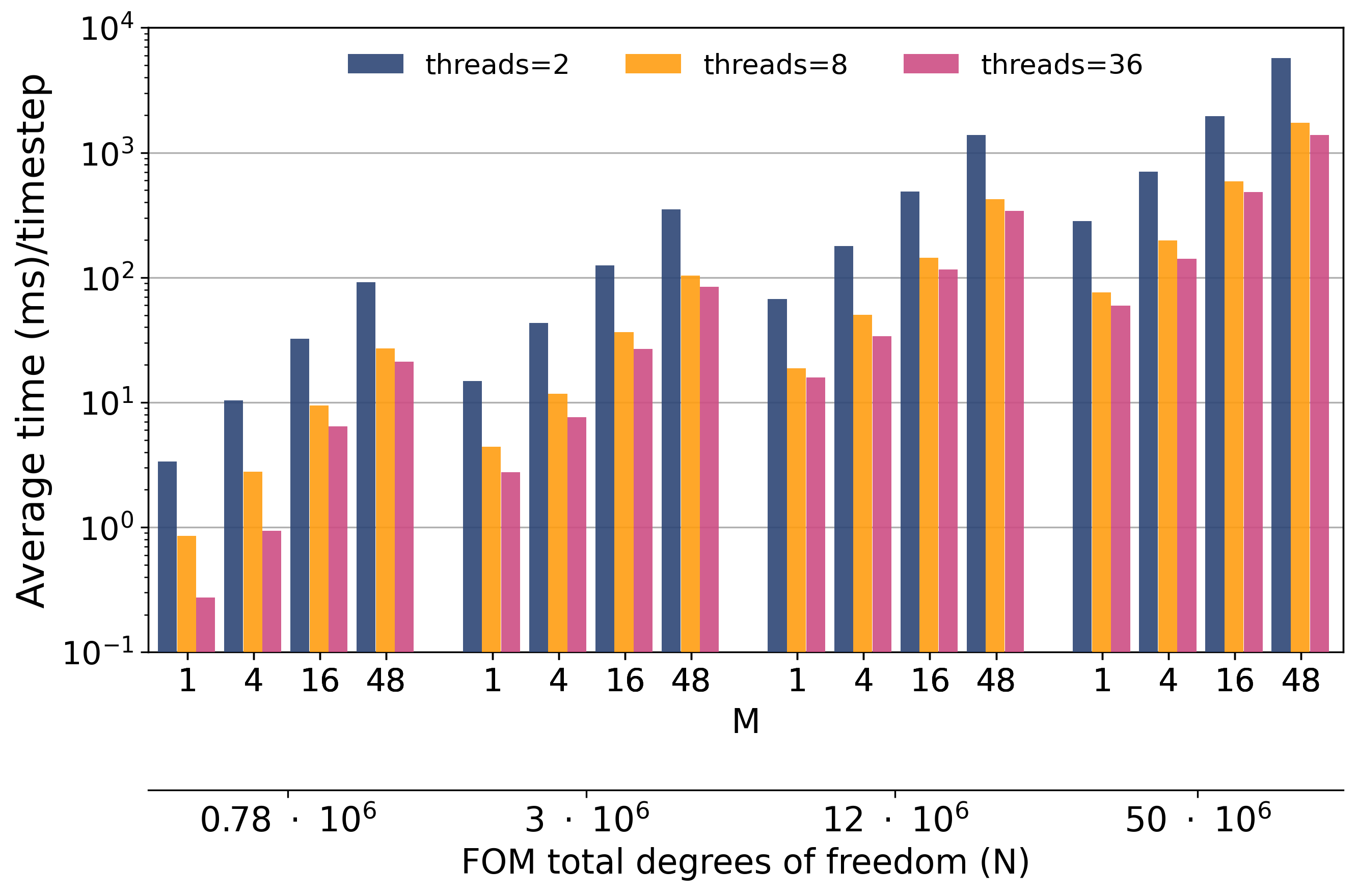}(c)
  \caption{Performance results obtained for the full-order problem
    in Eq.~\eqref{eq:shwave_ode_tensor} showing (a) the computational
    throughput (GFlops), (b) the memory bandwidth (GB/s),
    and (c) the average time (milliseconds) per time step,
    for various thread counts, problem and forcing sizes $\nRuns$.
    The limits of the vertical axes are the same as those used
    in Figure~\ref{fig:romScaling}.}
\label{fig:fomScaling}
\end{figure}

Figure~\ref{fig:fomScaling} shows the computational throughput (GigaFLOPS) in panel~(a),
the memory bandwidth (GB/s) in panel~(b) and
the average\footnote{The average time per step is computed by timing the full simulation and dividing
  by the total number of steps, which we set in all cases to $1000$.}
time (milliseconds) per time step in panel~(c)
for a representative subset of values of thread counts and $\nRuns$.
We make the following observations.
First, Figures~\ref{fig:fomScaling}~(a,b) show that as the problem size increases
and the thread count increases to $36$ the computational throughput
plateaus around $10$~GFlops (which is in line with other typical sparse
kernels~\cite{li2015}) and the memory bandwidth around about $68$~(GB/s),
which is close to the max bandwidth ($\sim 85$~GB/s)
of the machine in its current configuration.
This indicates a good performance, and confirms that the problem
is memory bandwidth bound.
Second, panel~(b) shows that for the smallest problem size ($\fomDim=0.78 \cdot 10^6$),
the memory bandwidth exceeds the theoretical one, indicating cache effects
playing a key role at that scale. These cache effects become increasingly less
evident as the problem size increases.
Third, if we fix the problem size, $\fomDim$, and thread count, we observe
that the performance improves if we use the rank-2 implementation.
This is because of the higher arithmetic intensity
of the \code{spmm} kernel in the \rtwofom~compared to \code{spmv}
in the \ronefom. For the sake of the argument, consider the problem size
$N=12\cdot10^6$ and the case with $36$ threads: Figure~\ref{fig:fomScaling}~(c) shows
that using $\nRuns=16$ allows us to simultaneously compute sixteen trajectories
for only a seven-fold increase in the iteration time with respect to $\nRuns=1$.
The plots also seem to suggest that the benefit of simulating multiple
trajectories at the same time, i.e., using $\nRuns \geq 2$, is evident
when going from $\nRuns=1$ to $\nRuns=4$, but then plateaus,
suggesting that there is a limiting, problem-dependent value of
$\nRuns$ after which no major gain is obtained.
Fourth, panel~(c) allows us to assess the strong scaling
behavior of the FOM problem. If we fix $\fomDim$ and $\nRuns$,
we observe a good scaling from 2 to 8 threads, but a degradation as we go from 8 to 36.
This is expected and explained as follows: since the problem
is memory bandwidth bound,
and 8 threads are nearly enough for the computational kernels to saturate the memory access
(see panel~(b)), one cannot expect a substantial improvement
in the performance if we increase the thread count.

\subsection{ROM performance and scaling} \label{sec:romScaling}

This section presents scaling and performance results
for the rank-1 and rank-2 Galerkin ROMs introduced in Eq.\eqref{eq:shwave_rom_ode_tensor}.
To carry out this analysis, we make the following simplifications:
(a) we use random data to fill the reduced operators in Eq.\eqref{eq:shwave_rom_ode_tensor}
(a demonstration and discussion of the Galerkin ROM accuracy on a real
shear wave problem is presented in \S~\ref{sec:romAccuracy} and \S~\ref{sec:mqRom});
(b) we turn off all input/output related to saving data and only collect timing information;
and (c) we use the same ROM size, $\romDim$, for the velocity and stresses generalized coordinates,
i.e., $\romDim_{\vp} = \romDim_{\stresses} = \romDim$.

We consider the following ROM sizes
$K \in \{256, 512, 1024, 2048, 4096\}$,
thread counts ${1,2,4,8,12,18,36,72}$, and
$\nRuns \in \{1,2,4,8,16,32,64,128,256,512,1024\}$.
Recall that the case $\nRuns=1$ corresponds to the \ronerom~while
$\nRuns \geq 2$ corresponds to \rtworom.
We remark that these choices of $\nRuns$, despite seemingly large,
are all feasible for the ROM problem thanks to the small state dimensionality.
To give an example, let us consider a Galerkin ROM problem of the form
Eq.\eqref{eq:shwave_rom_ode_tensor} with the largest size, i.e., $K=4096$,
and suppose we collect a total of $500$ reduced states.
If we use $M=1000$, which corresponds to simulating $1000$ forcing
realizations simultaneously, and considering we have one equation
for the velocity and one for stresses,
it would require only $\approx 65$~MB for the generalized coordinates,
$\approx 268$~MB for the reduced system matrices and $\approx 33$~GB to store
all the state snapshots.
This is well within the memory capacity of modern computing nodes.
Note that we are not accounting for the size of the basis because
the reduced operators can all be precomputed offline.

For both \ronerom{} and \rtworom{} we use a column-major layout
for all the operators; this is a suitable to interface to the
BLAS library used by the kokkos-kernels to execute all the dense kernels. 
All runs are completed using the following OpenMP
affinity\footnote{\url{https://www.openmp.org/spec-html/5.0/openmp.html}}:
\code{OMP\_PLACES=cores} and \code{OMP\_PROC\_BIND=true}.
\begin{figure}[!t]
  \centering
  \includegraphics[width=0.675\textwidth]{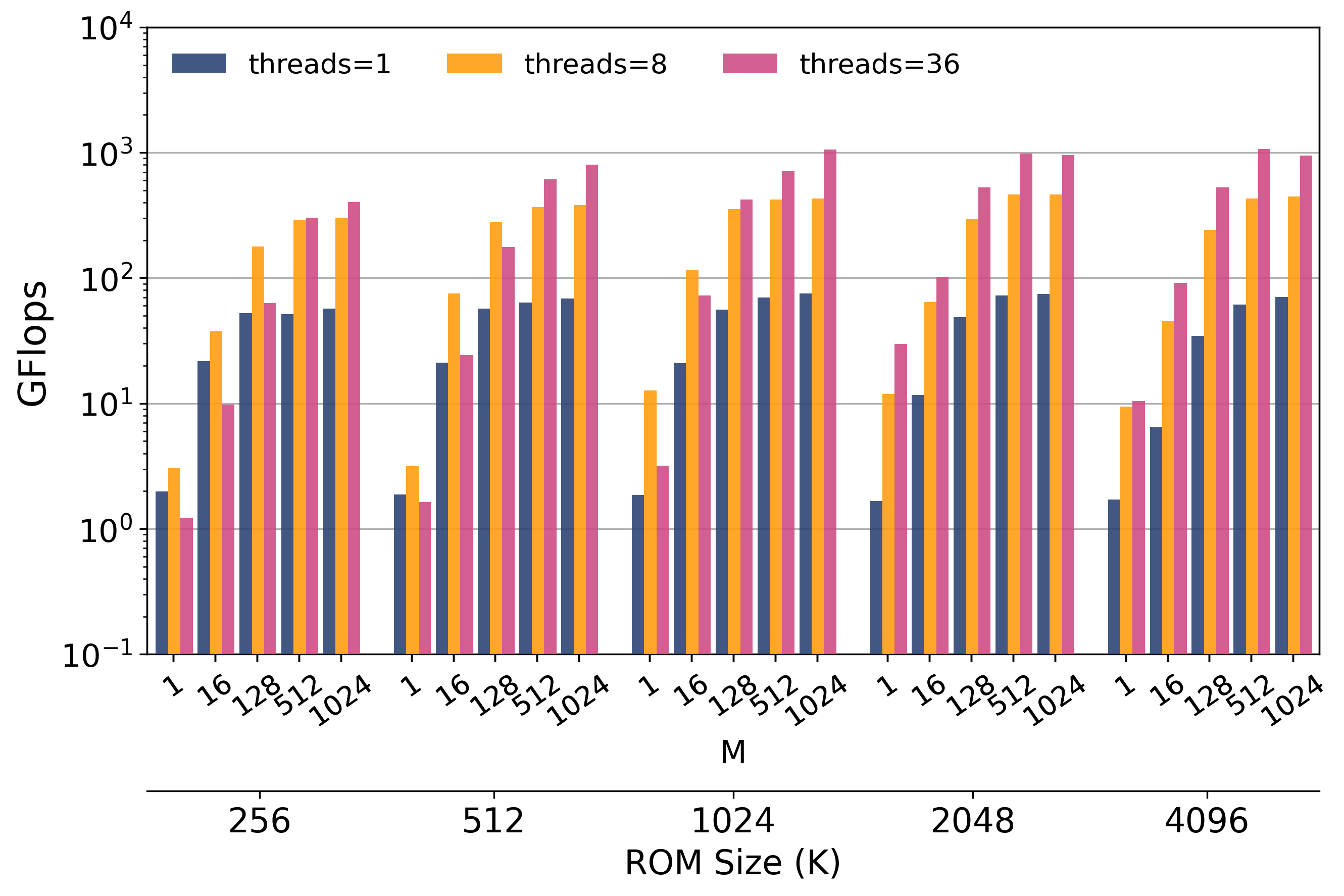}\\
  \includegraphics[width=0.675\textwidth]{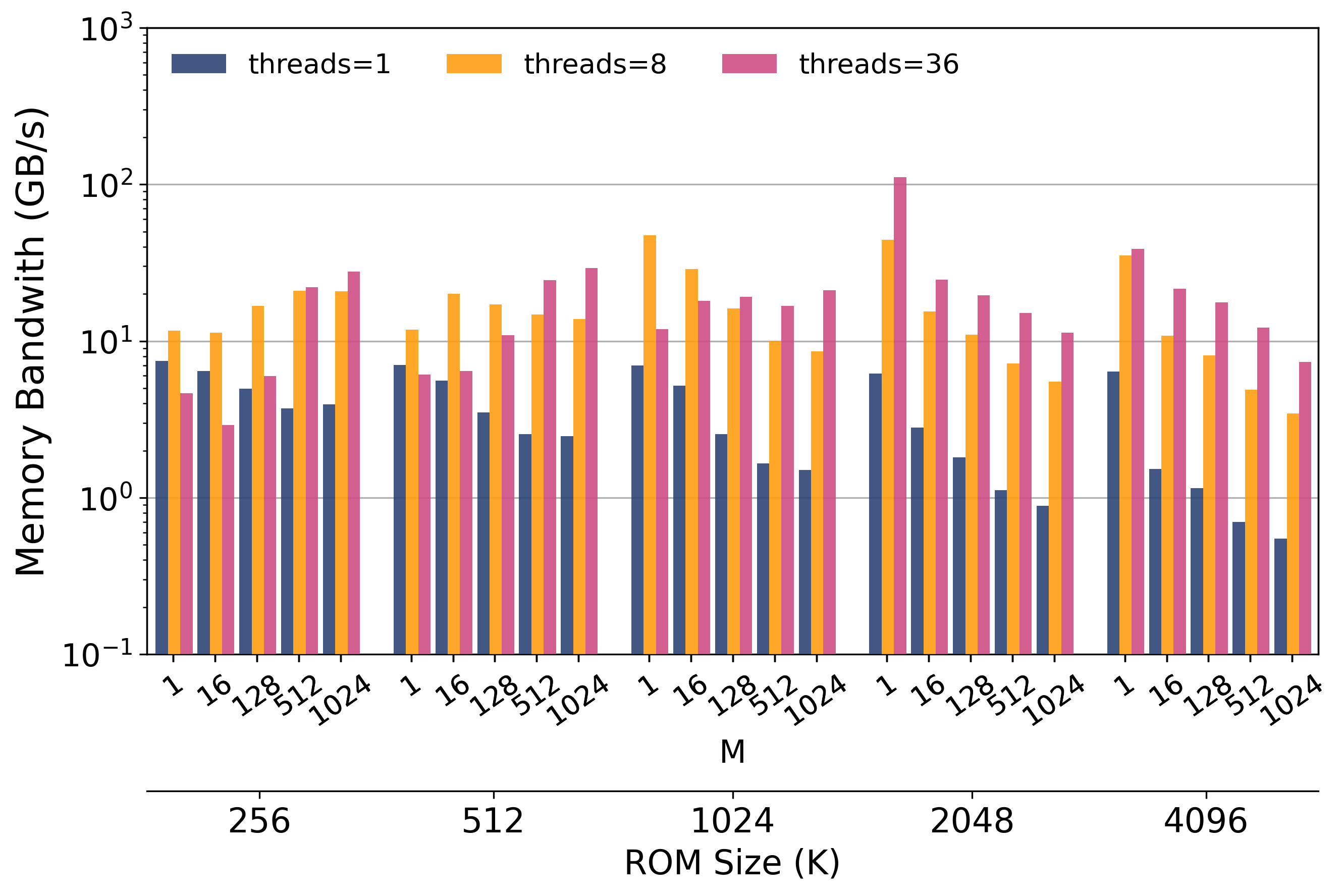}\\
  \includegraphics[width=0.675\textwidth]{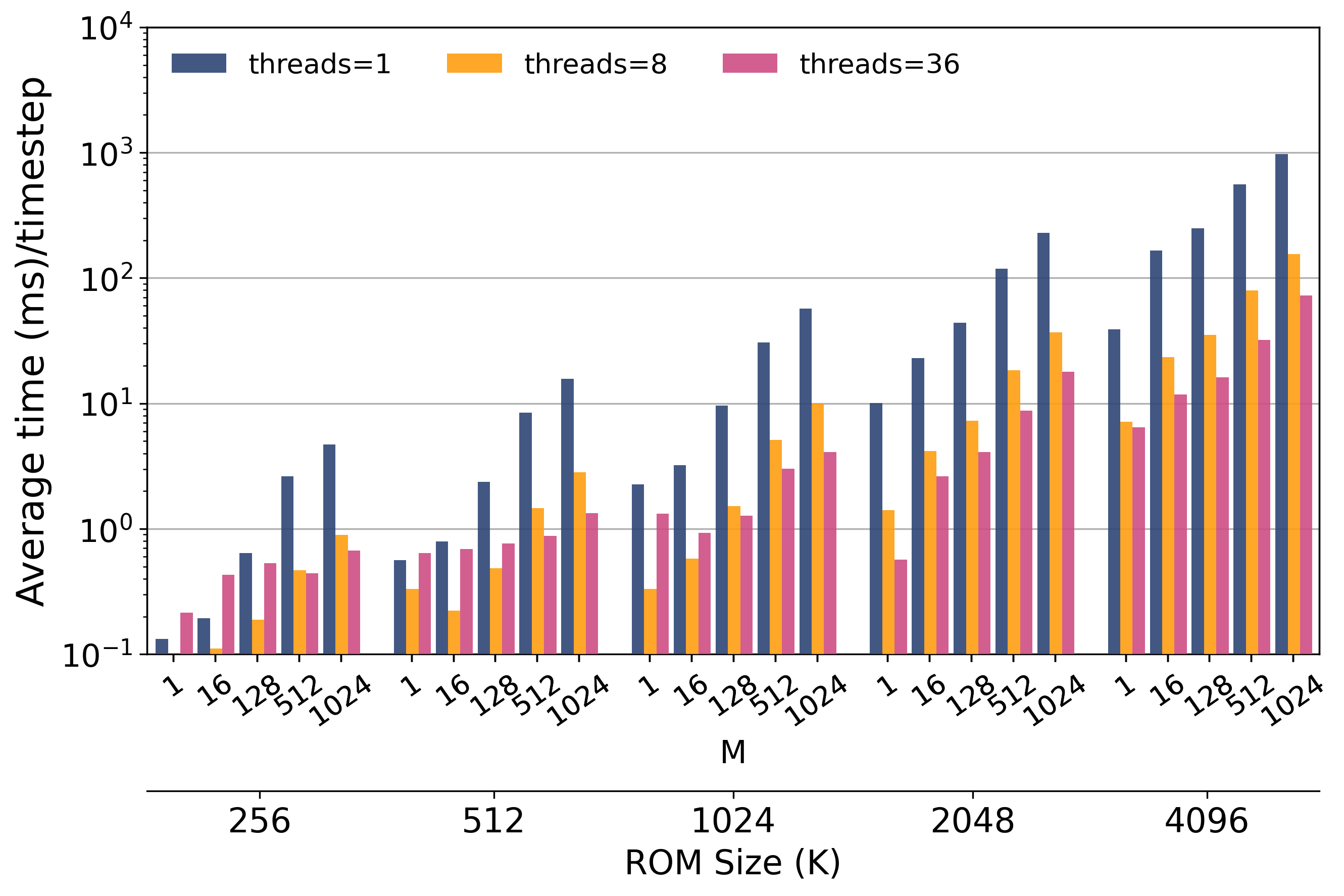}
  \caption{Performance results obtained for the ROM problem
    in Eq.~\eqref{eq:shwave_rom_ode_tensor} showing the (a) computational
  throughput (GFlops), (b) memory bandwidth (GB/s),
  and (c) average time (milliseconds) per step, for various thread counts,
  forcing size $\nRuns$, and number of modes $\romDim$.
  %The limits of the vertical axes are the same as those used
  %in Figure~\ref{fig:fomScaling} to ease a visual comparison.
  }
  \label{fig:romScaling}
\end{figure}

Figure~\ref{fig:romScaling} shows the computational throughput (GigaFLOPS) in panel~(a),
the memory bandwidth (GB/s) in panel~(b) and the
average\footnote{The average time per step is computed by timing the full simulation and dividing
  by the total number of steps, which we set in all cases to $4000$.}
time (milliseconds) per time step in panel~(c) for a representative
subset of values of threads and $\nRuns$.
We make the following observations.
First, Figures~\ref{fig:romScaling}~(a,b) clearly show that
\ronerom~(i.e. $\nRuns=1$) is a memory bandwidth-bound problem,
while \rtworom~($\nRuns>1$) is a compute-bound problem.
This is evident because the case $\nRuns=1$
yields a limited throughput, namely $\ordOf{10}$~Gflops,
which remains substantially unchanged if we increase the ROM size $\romDim$.
Also, if we fix $\nRuns=1$ and increase the number of threads, we observe
a minor improvement only up to 8 threads, which is already enough to saturate the system.
On the contrary, when $\nRuns \ge 16$, using more threads is generally
beneficial and the throughput reaches a maximum of 1TFlops when the cores are fully utilized.
Compared to the GFlops obtained for $\nRuns=1$, the one for $\nRuns=16$ is
one order of magnitude larger, and increases to two orders of magnitude for $\nRuns \ge 512$.
Second, for the ROM sizes explored here, the computational throughput
achieves its maximum when $\nRuns=512$ and no major improvement is obtained
using $\nRuns=1024$. For example, for $\romDim=2048$,
using $\nRuns=512$ and 36 threads yields about 1TFlops,
which remains the same for $\nRuns=1024$.
Third, panel~(c) allows us to assess the strong scaling behavior.
We observe excellent scaling when $\nRuns \ge 16$ and the ROM size
is sufficiently large, while a poor scaling is generally obtained when $\nRuns=1$,
which is a direct consequence of the different nature of the kernels.

\subsection{When should an analyst prefer the \rtworom?} \label{sec:romSpeedup}
The results above highlight the excellent
performance of \rtworom{} and quantified the main differences between it and \ronerom.
A natural question arises: when should an analyst prefer \rtworom{} over \ronerom?
Obviously, this question only makes sense in the context of a
many-query study where the interest is in simulating an ensemble
of trajectories corresponding to multiple realizations of the forcing function,
as in a typical forward propagation study in UQ.

Suppose we are given a target ROM size, $K$, and need to simulate an ensemble
of $P$ trajectories (where $P$ is large, $P \gg 10$) from realizations of the forcing term,
while needing to meet these two constraints:
(a) we have a limited budget of cores available, e.g., a single node
with $36$ physical cores; (b) due to, e.g., memory constraints, $1024$ is the
maximum feasible number of trajectories that can simultaneously live on the node.
These constraints are arbitrarily set here for the sake of the argument,
but are reasonable values nonetheless.
What combination of thread count, $n$, and
number of simultaneous trajectories, $\nRuns$, would be the most efficient
to obtain those $P$ samples while satisfying the given constraints?
For example, one could launch $36$ single-threaded runs in parallel,
each using $\nRuns=1$, and then repeat until all $P$ runs are completed.
This implies that, at any time, all $36$ cores of the node would be occupied,
and $36$ realizations of the forcing term would be running simultaneously,
which means that both the core budget and the memory constraint are satisfied.
A minor variation would be to run $18$ two-threaded runs at the same
time each using $\nRuns=1$. This would still satisfy both the core
budget and the memory constraint. The most interesting scenarios
arise when we vary $\nRuns$.
Generalizing, this is a (discrete) constrained optimization problem,
since we need to optimize over number of threads and $\nRuns$.
%and possibly over heterogenous combinations of $\nRuns$ and threads.
Solving this in a general context is outside the scope of this work,
but we provide the following insights.
%% Note that the memory constraint here is more relaxed than the one in the FOM analysis,
%% because due to the small size of the ROM problem, we can affor to run many more
%% forcing samples at the same time than we did in the FOM.

Let $\tau(\romDim, n, \nRuns)$ represent the runtime to complete a {\it single}
Galerkin ROM simulation of the form Eq.\eqref{eq:shwave_rom_ode_tensor}
with ROM size $\romDim_{\vp} = \romDim_{\stresses} = \romDim$,
using $n$ threads and a given value $\nRuns$.
It follows that the total runtime to complete trajectories for
$P$ forcing realizations with a budget of $36$ threads can be expressed as
\begin{equation}
\tau^{P}(\romDim, n, \nRuns) = \tau(\romDim, n, \nRuns) \frac{P}{\frac{36}{n} \nRuns},
\end{equation}
because $\frac{36}{n}$ is the number of independent runs executing
in parallel on the node with each run responsible of
computing $\nRuns$ trajectories.
We can define the following metric
%the quantify how advanatageous would be to use a \rtworom~with $\nRuns>=2$ over \ronerom~with
\begin{equation} \label{eq:romSpeedup}
  s(\romDim,n,\nRuns)
  = \frac{\tau^P(\romDim,1,1)}{\tau^P(\romDim, n, \nRuns)}
  = \frac{\tau(\romDim,1,1)}{\tau(\romDim, n, \nRuns)} \frac{\nRuns}{n},
\end{equation}
where $s(\romDim,n,\nRuns)>1$ indicates \rtworom~is more efficient
than \ronerom, while the opposite is true for $s(\romDim,n,\nRuns)<1$.
This metric can be interpreted as a speedup (or slowdown) factor.
\begin{figure}[t]
  \centering
  \includegraphics[trim=0 0 0 0,clip,width=0.47\textwidth]
                  {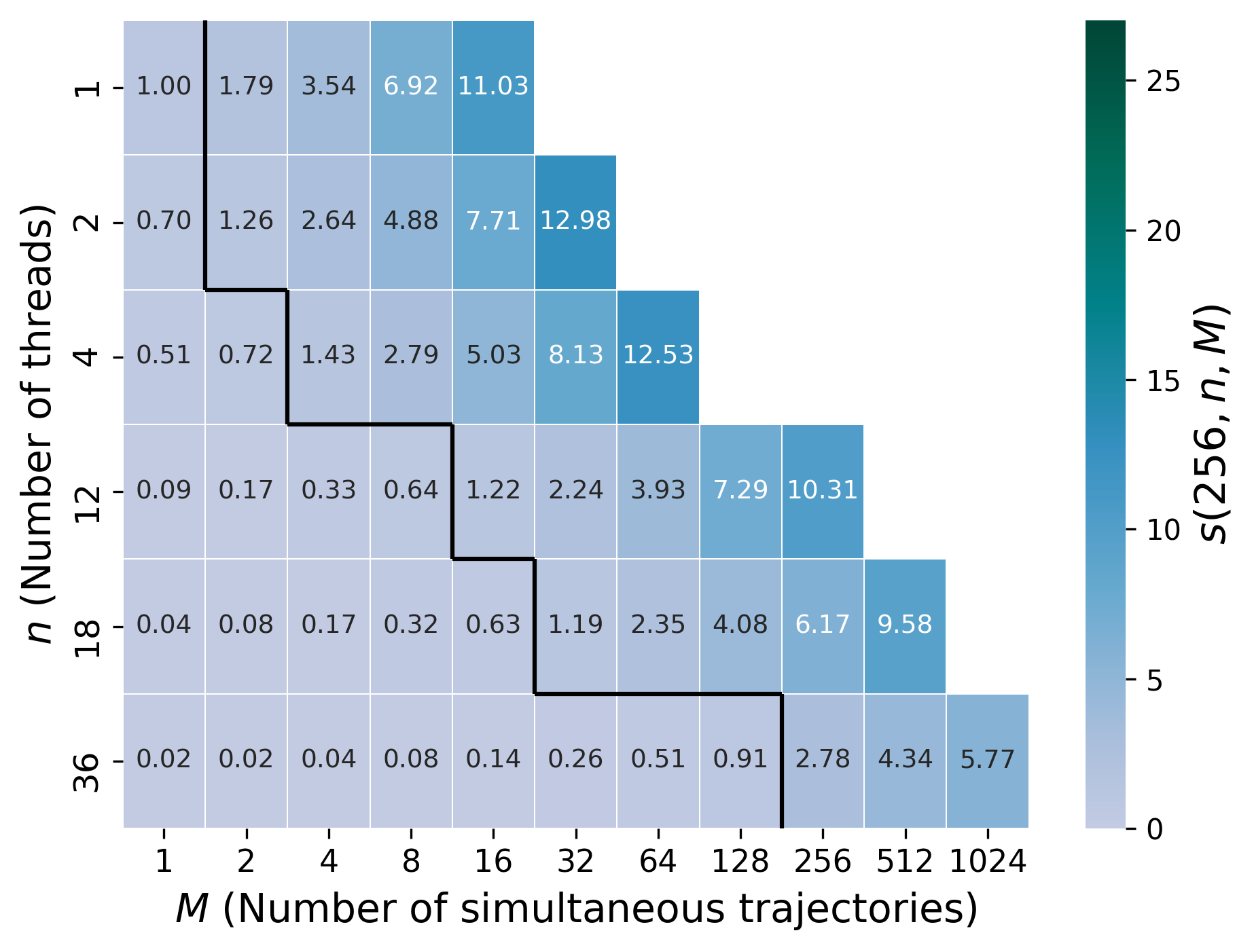}(a)
  \includegraphics[trim=45 0 0 0,clip,width=0.43\textwidth]
                  {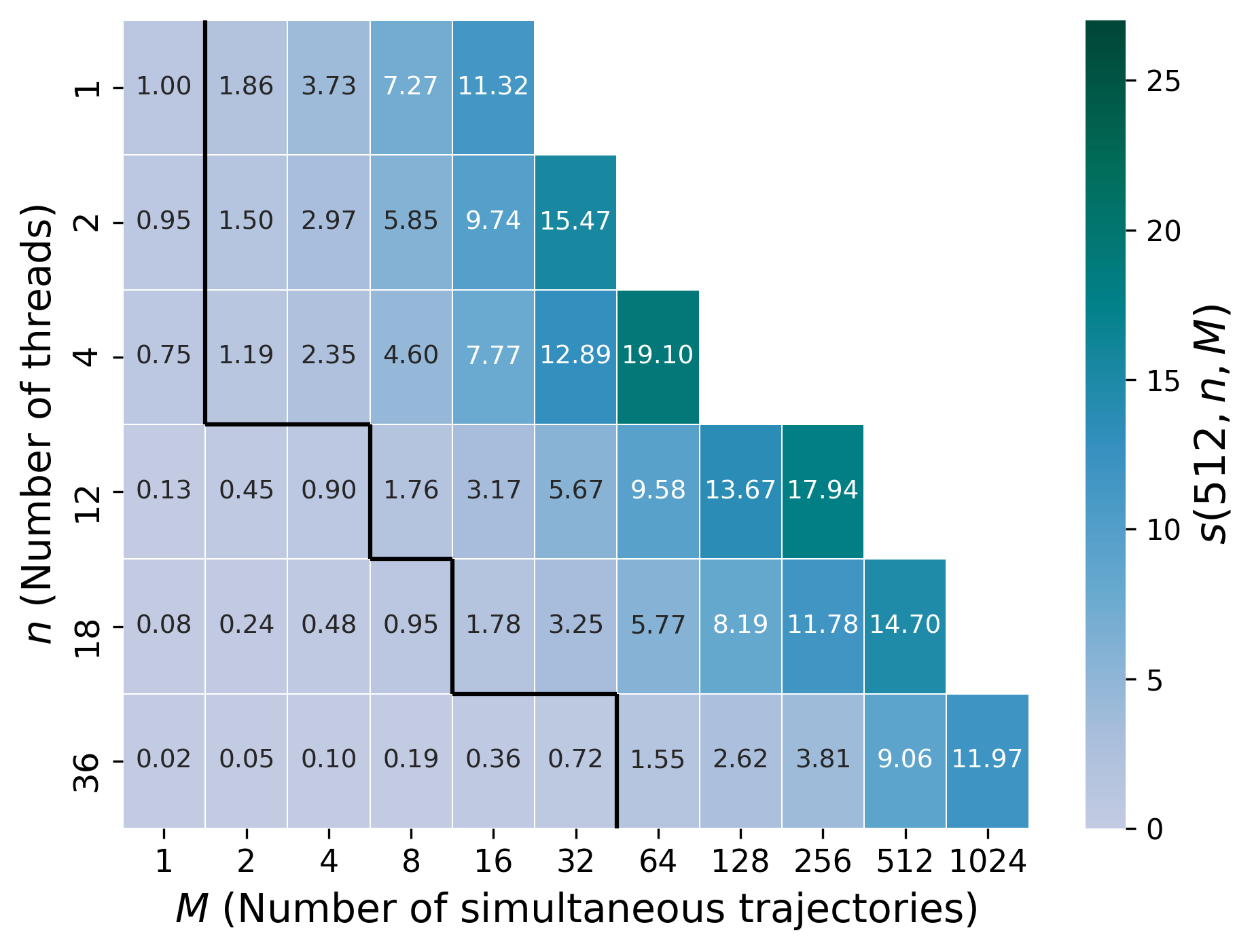}(b)\\
  \centering
  \includegraphics[trim=0 0 0 0,clip,width=0.47\textwidth]
                  {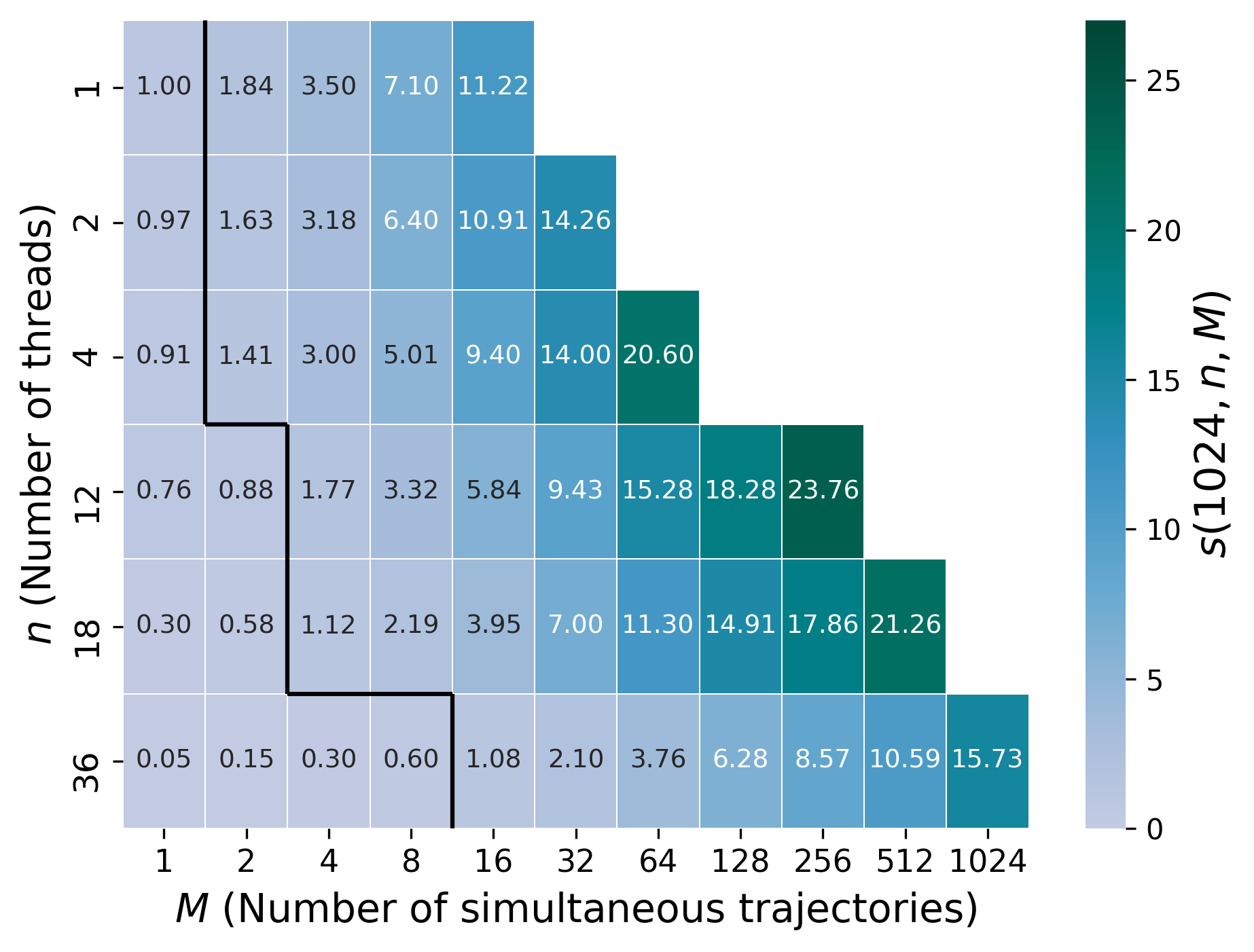}(c)
  \includegraphics[trim=45 0 0 0,clip,width=0.43\textwidth]
                  {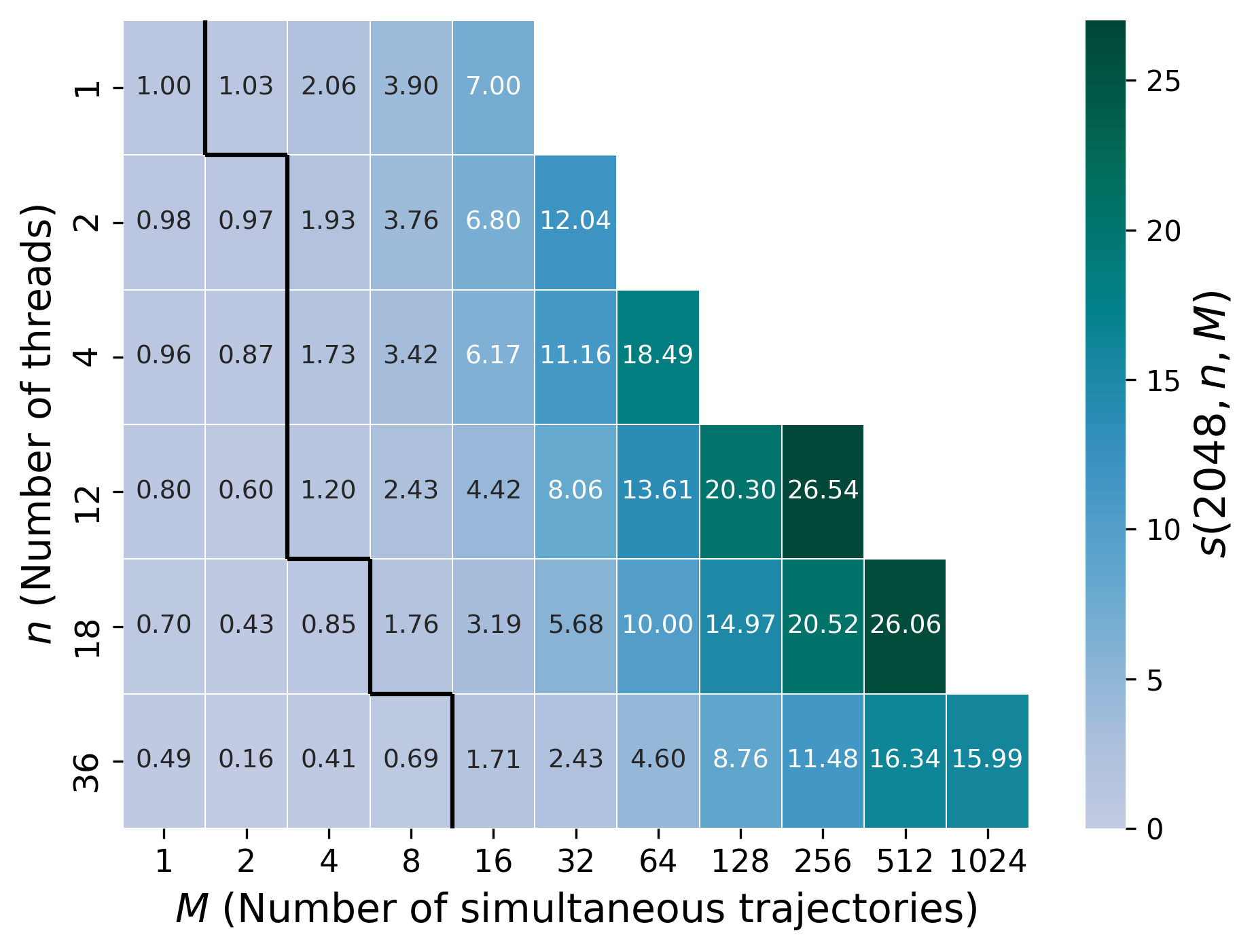}(d)
  \caption{Heatmap visualization of $s(\romDim, n, \nRuns)$ (see Eq.~\ref{eq:romSpeedup})
    computed for ROM sizes ${\romDim=256}$~(a), $512$~(b), $1024$~(c) and $2048$~(d),
    and various values of $\nRuns$ and $n$}. On each plot, the solid black line separates the
    cases where $s(\romDim, n, \nRuns)>1$ (i.e., where \rtworom~is more convenient)
    from those where $s(\romDim, n, \nRuns)<1$ (i.e., where \ronerom~is more convenient).
    The white region in each plot identifies the non-admissible cases, i.e.,
    combinations violating the constraints listed in \S~\ref{sec:romSpeedup}.
\label{fig:romSpeedup}
\end{figure}

Figure~\ref{fig:romSpeedup} shows a heatmap visualization of
$s(\romDim, n, \nRuns)$ computed for various values of $\nRuns$ and $n$, and
ROM sizes $\romDim \in \{256, 512, 1024, 2048 \}$. To generate these
plots, we used the runtimes obtained in \S~\ref{sec:romScaling}.
The plots allow us to reason about what is the most efficient setting.
For example, looking at Figure~\ref{fig:romSpeedup}~(a),
which corresponds to $\romDim=256$, we observe that using
\rtworom{} with $M=32$ and $n=2$ yields $s(256, 2, 32) = 12.98$,
which means that for this setup the \rtworom{} is about $13$ times
more efficient than using \ronerom~with $n=1$.
Depending on the given ROM size, similar conclusions can be made.
The second key observation is that as the ROM size increases,
it becomes increasingly more convenient to rely on \rtworom.
For example, if we fix $n=12$ and $M=256$, we see that when $\romDim=256$ we
have a 10X speedup with respect to \ronerom~with $n=1$
(see Figure~\ref{fig:romSpeedup}~(a)), but we reach a 26X speedup
for $\romDim=2048$, see Figure~\ref{fig:romSpeedup}~(d).

%We conclude this section by partially answering the question
%posed at the beginning: when should an analyst prefer \rtworom~over the \ronerom?
% \begin{fmytheo}
The above results suggest that, if the goal is to compute an ensemble
of trajectories for different realizations of the forcing term
and the main cost function is the runtime, \rtworom~should always
be preferred over the \ronerom.
% \end{fmytheo}

\subsection{Predictive and reproductive ROM accuracy} \label{sec:romAccuracy}

We now discuss the {\it accuracy} of the Galerkin ROM for the elastic
shear wave test case under consideration.
We consider the scenario where the period, $T$, parameterizing the
forcing signal varies over a fixed interval, $T \in [31, 69]$,
and aim to quantify how accurately the Galerkin ROM can approximate
the FOM predictions as $T$ varies over that interval.
Note that since here the focus is on quantifying the {\it accuracy},
it does not matter which ROM formulation is used.
We use \rtworom{} to perform the analysis because
it is more efficient, but the same results would be obtained using \ronerom{}.

The complete definition of the problem is as follows.
We use the PREM as material model, a total simulation time of
$\tfinal=2000$~(seconds) and a time step size $\Delta t = 0.1$~(seconds) to ensure stability.
The forcing acts at $150$~(km) depth, and is modeled as a Ricker wavelet
with $90$ seconds delay and period $T$. This delay, combined with the total
integration time, represents a system that is at rest for a small
period of time, and then excites and decays due to the action of the forcing.
%The forcing period is assumed to be $T \in (35,65)$~(seconds).
For the FOM runs, we use the grid with $\fomDim=3,143,168$ total degrees of freedom,
which we recall stems from using a $512 \times 2048$ velocity grid.
This grid is chosen because, given $T=31$ and the PREM material model,
it is the smallest one (in powers of two) satisfying the numerical
dispersion criterion we use in our code.\footnote{To ensure the numerical
dispersion is satisfied, for any given grid we require its maximum
grid spacing, $h$, to satisfy $h \leq T v_s^{min}/\lambda$ where $\lambda$
is the target number of grid points per wavelength which we set to $10$
and $v_s^{min}$ is the minimum shear velocity in the material model.}

To compute the POD modes for the ROM, we proceed as follows: we run the FOM
at $T=35,65$, save the velocity and stress states every $2$ seconds,
and then merge the data from both training points into a single snapshot
matrix which is used to compute the POD modes via SVD.
We deliberately choose only two training points
for two main reasons: (a) it is the smallest training set one can choose
to sample a one-dimensional parameter space; and (b) it mimics
a realistic situation where limited data are available.
This occurs, for example, when
a computational model is very expensive to run such that an analyst can
only afford a few runs, or where experimental data is collected
at just a few parameter points and more
experiments are not affordable. For such cases, in fact, relying on
data-driven models requiring large training datasets is unfeasible.
We thus believe it is interesting to assess whether
one can construct a sufficiently accurate ROM given limited data,
which, if positive, would highlight the full power of ROMs versus
purely data-driven models.

Our goal is to characterize both a reproductive and predictive scenario.
A reproductive scenario occurs when the ROM accuracy is quantified and
evaluated at the same (training) points in the parameter
space used to generate the basis functions of the ROM.
In this case, this means evaluating the ROM accuracy for $T \in \{35,65\}$.
A predictive scenario is where the ROM accuracy is evaluated
at parameter instances (test points) that differ from those used to generate
the basis functions of the ROM. Here, the predictive analysis is run
at $10$ random samples over the parameter range $(35, 65)$,
as well as $T=31$ and $T=69$, {\it outside} the training range.
We remark that this is a key choice because it allows us to assess
the extrapolation capabilities of the ROM, which generally is one of
the main weaknesses of purely data-driven methods.

To quantify the accuracy of the ROM, we rely on the following relative error
\begin{equation} \label{eq:errors}
  E_{\ell_2}
  = \frac{\normtwo{
      \state(\tfinal;\paramsA, \paramsF) -
      \approxState(\tfinal;\paramsA, \paramsF)}}
  {\normtwo{\state(\tfinal;\paramsA, \paramsF)}}
  %\qquad E_{\ell_{\infty}} = \frac{\norminf{x(\mu) - \tilde{x}(\mu)}}{\norminf{x(\mu)}}
\end{equation}
where $\state(\tfinal;\paramsA, \paramsF)$ is the FOM state (for either velocity
or stresses) computed at the final time $\tfinal$ for the given
parameters $\paramsA, \paramsF$, and $\approxState(\tfinal;\paramsA, \paramsF)$
is the state field reconstructed from the ROM results.
Relying on the final time is sufficient for the purposes of this section,
and in \S~\ref{sec:mqRom} we show some results quantifying the error
over the full simulation time.

\begin{figure}[!t]
  \centering
  \includegraphics[width=0.45\textwidth]{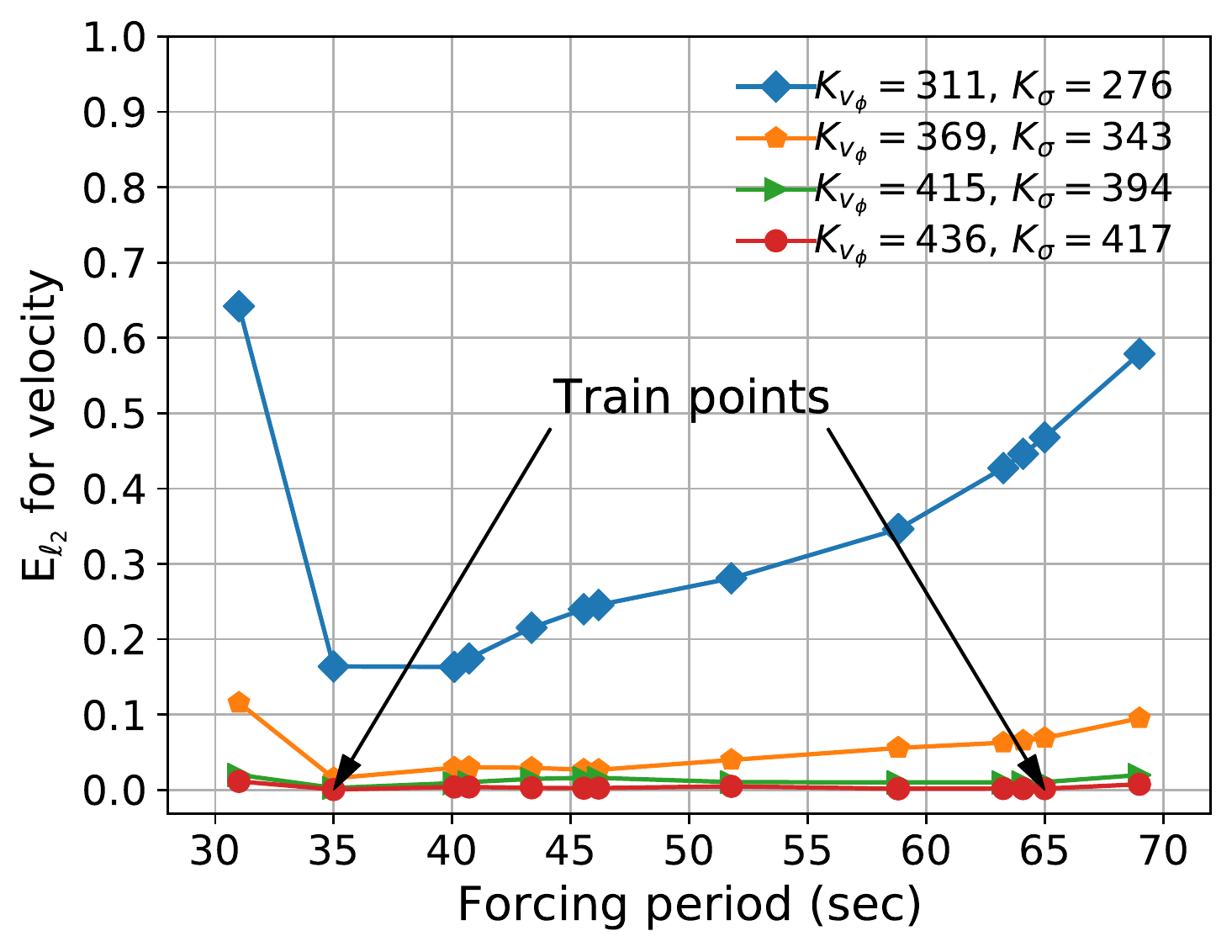}(a)
  \includegraphics[width=0.45\textwidth]{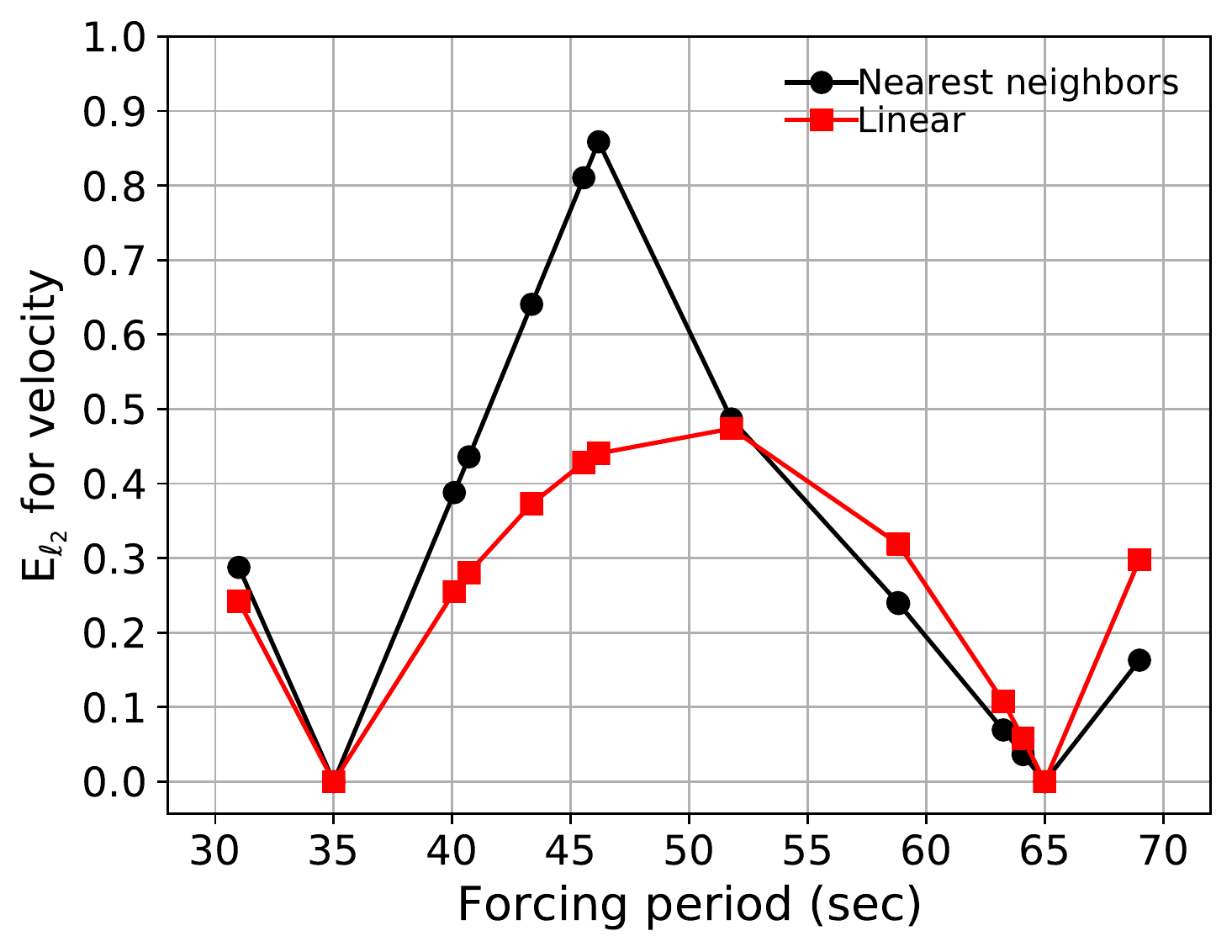}(b)\\
  \includegraphics[trim=65 0 112 0,clip,width=0.26\textwidth]{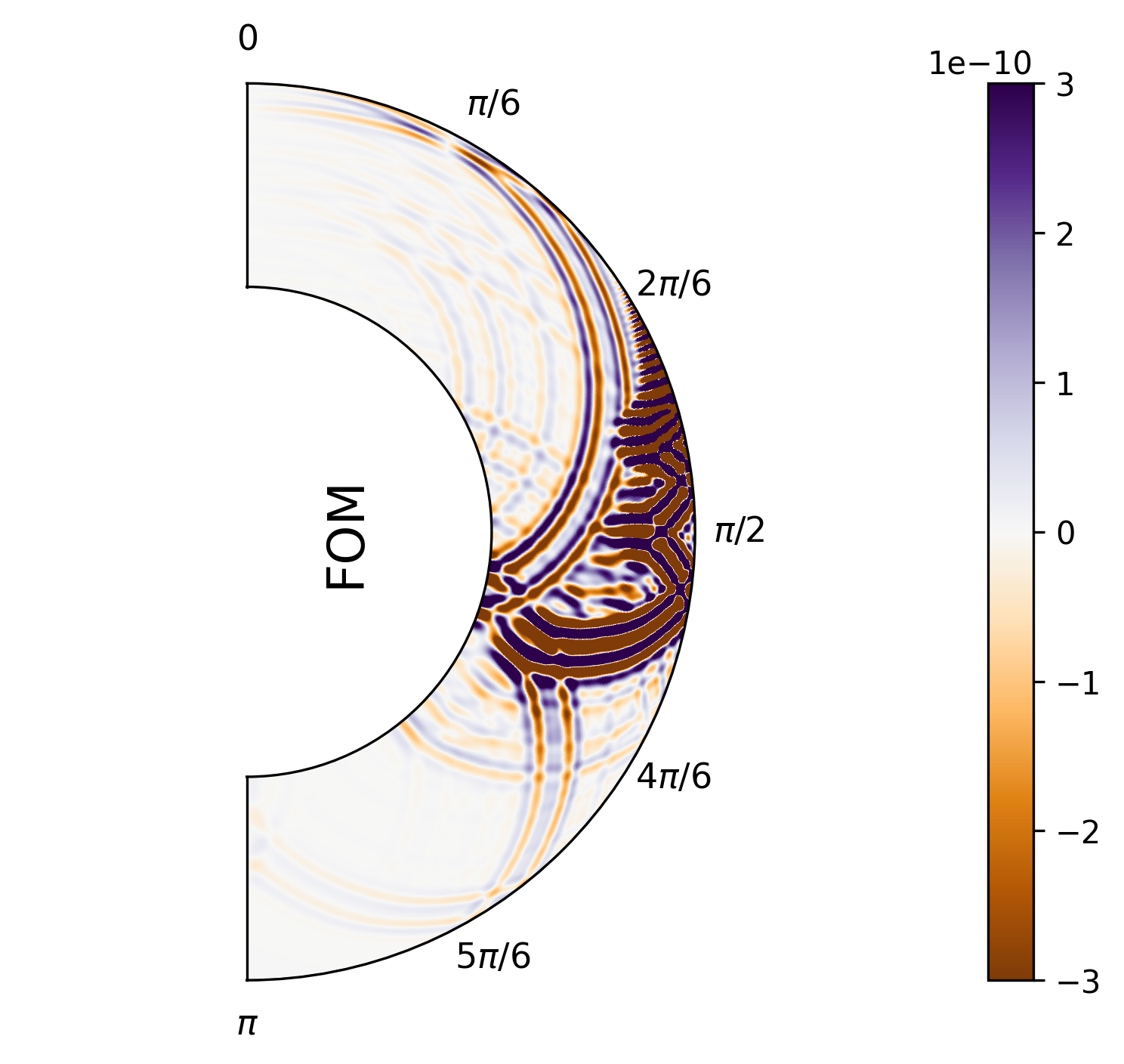}
  \includegraphics[trim=72 0 112 0,clip,width=0.25\textwidth]{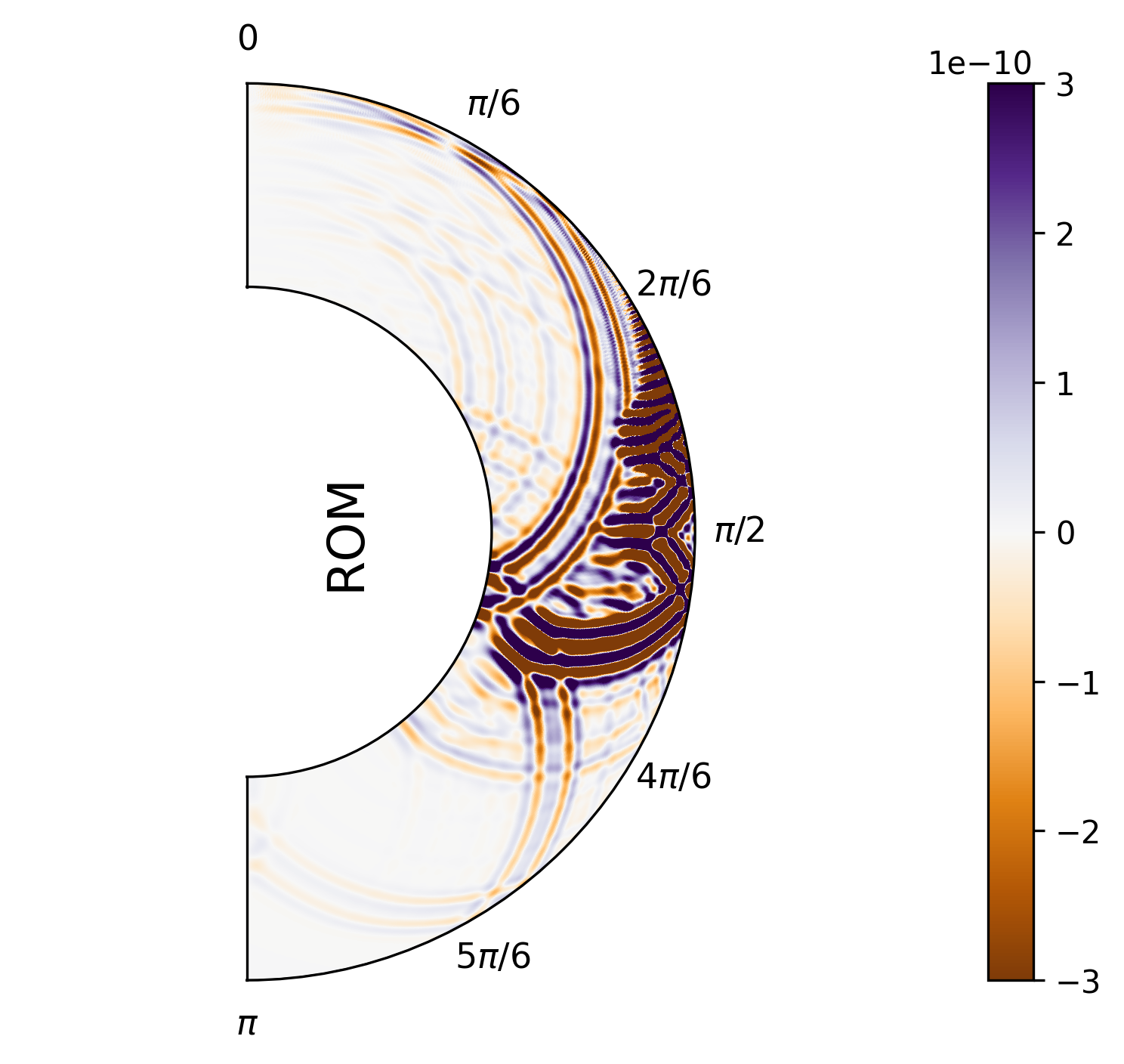}
  \includegraphics[trim=72 0 112 0,clip,width=0.25\textwidth]{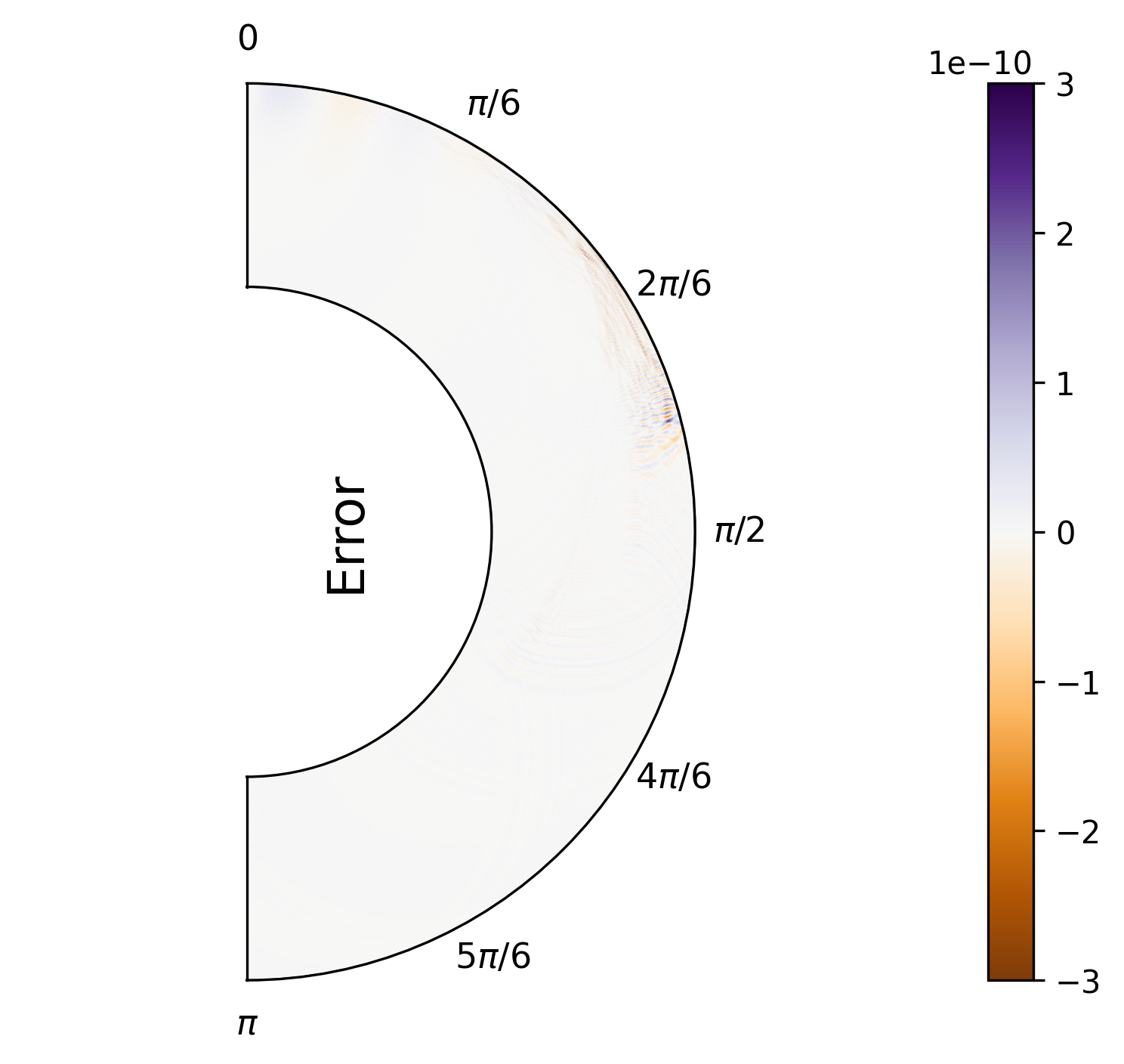}
  \includegraphics[trim=280 0 8 0,clip,width=0.075\textwidth]{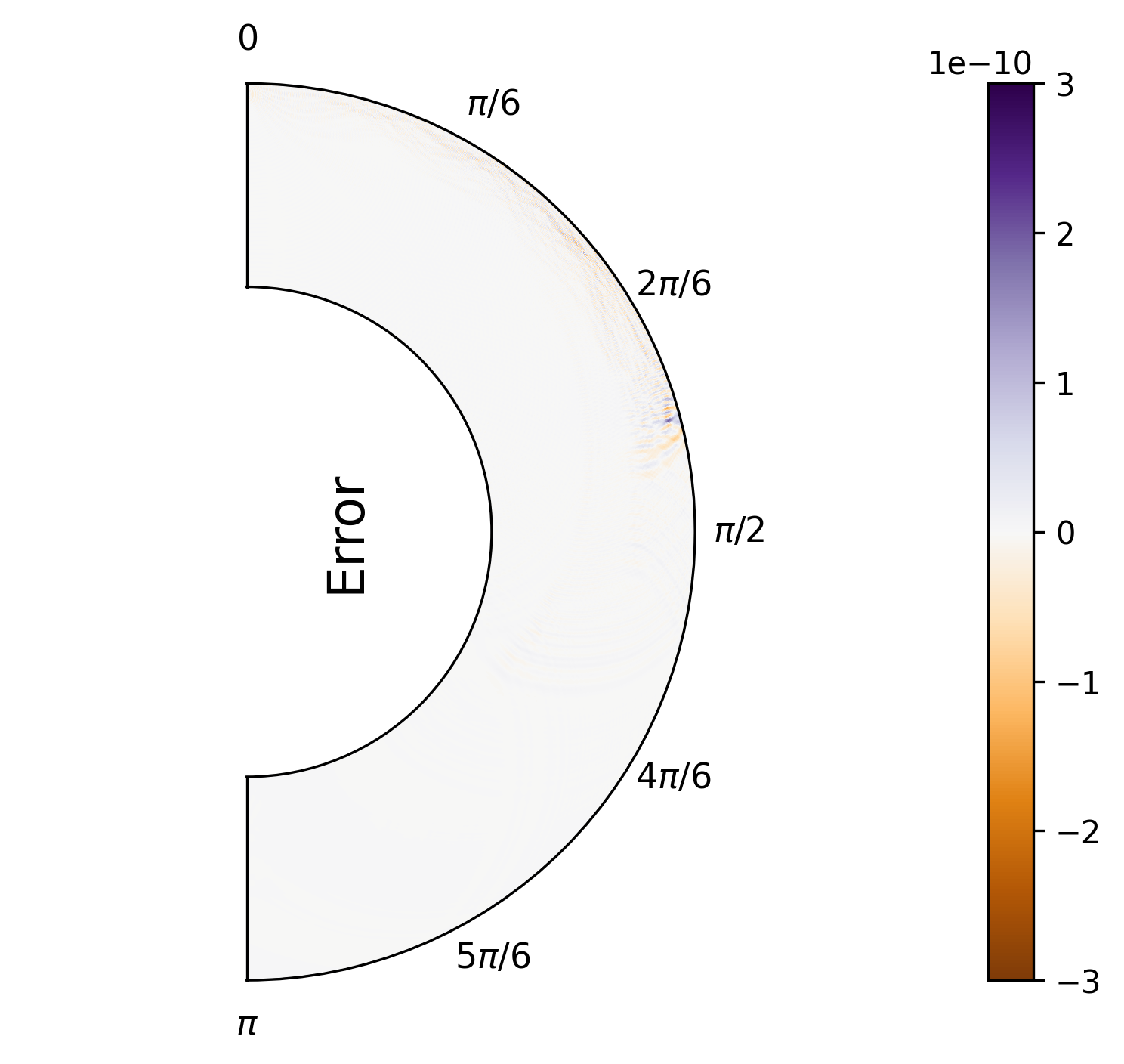}(c)
  \caption{Panel~(a) shows the relative error, see Eq.\eqref{eq:errors},
    computed for the velocity degrees of
    freedom between the FOM solution and the corresponding ROM-based
    approximation as a function of the period $T$ and for various ROM sizes $K$.
    Panel~(b) shows the same relative error metric computed between
    the FOM and the state approximation obtained via nearest neighbor and linear
    interpolation using the FOM state at $T=35$ and $T=65$ as training points.
    Panel~(c) shows the velocity wave field computed with the FOM (left), the ROM
    with $(436,417)$ modes (middle), as well as their error field (right) at
    $\tfinal=2000$~(seconds) for the extrapolation point $T=69$.}
\label{fig:romerrorsAcc2}
\end{figure}

Figure~\ref{fig:romerrorsAcc2}~(a) shows the relative error \eqref{eq:errors}
computed for the velocity degrees of freedom as a function
of the period $T$. The results are shown for $(\romDim_{\vp}, \romDim_{\stresses})
\in \{(311, 276), (369, 343), (415, 394), (436, 417)\}$,
which corresponds to truncating $10^{-5}$, $10^{-7}$, $10^{-9}$ and $10^{-10}$ 
of the POD cumulative energy, respectively. 
To compute the cumulative energy, we use the same fomula defined in~\cite{swischuk2020}).
%$99.999 \%$, $99.99999 \%$, $99.9999999 \%$, and $99.99999999 \%$
As expected, increasing the ROM size increases the accuracy.
We observe that to obtain errors smaller than $5 \%$ {\it within} the training interval
one would need to use at least $(\romDim_{\vp},\romDim_{\stresses})=(369,343)$,
but that this choice would still yield a large error of about $12 \%$
at the extrapolation point $T=31$. To have small errors both within
the training interval and in the extrapolation regime, one would need
at least $(\romDim_{\vp},\romDim_{\stresses})=(415,394)$.
If a sufficient number of modes is chosen, the accuracy of the ROM
is excellent across the full range and also
at the extrapolation points. Similar error results are obtained for
the stresses but we omit the corresponding plot for brevity.
We remark that we purposefully omit the data obtained for lower energies
because those cases yield unacceptable results reaching relative
errors that are above 100 \% for some values of $T$.
For this application, the convective nature of the problem can
force the need for a large number of modes
to accurately predict the dynamics.
However, this is not a weakness but can be an
advantage in this case because, as discussed in \S~\ref{sec:romScaling},
the compute-bound kernels involved in \rtworom{}
become more efficient and scalable
as $\romDim$ increases (up to value which is
typically large, $\ordOf{10000}$).

For the sake of argument, Figure~\ref{fig:romerrorsAcc2}~(b) shows
the same relative error metric computed between the FOM and the state
obtained via nearest-neighbor and linear interpolation
using the FOM states at $T=35$ and $T=65$ to compute the approximation.
A linear interpolation is the best we can do with two data points.
Figure~\ref{fig:romerrorsAcc2}~(b) shows that both nearest-neighbor
and linear interpolation have quite poor accuracy overall,
with errors reaching up to $90\%$ inside the training range and
consistently larger than $15\%$ at the extrapolation points.
This indicates that, for this test case, the ROM is much more
robust and accurate than interpolation.

Panel~(c) shows a contour plot of the wave field
at the final time $\tfinal=2000$~(seconds) for the source signal
period $T=69$ (recall that this is an extrapolation point) computed with
the FOM (left), ROM with ${(\romDim_{\vp},\romDim_{\stresses})=(436,417)}$ (middle),
and their error field (right).
We observe very little difference between the FOM and ROM results,
mostly near the earth surface in the range $2\pi/6 < \theta < \pi/2$.

Given the scope of this work, the above discussion was limited
to a representative analysis, but we anticipate a full
comparison between the FOM, ROM and other surrogate models
(e.g., polynomials or neural nets) to be an interesting study to perform.
Several questions arise:
\begin{itemize}
\item How many more training points would one need to use for a polynomial
  approximation to match the ROM accuracy?
\item What is the best performing surrogate when the number
  of training points is larger than two?
\item What is the trade off between the cost of computing the surrogate and its accuracy?
\item How would the results change if one considered a higher-dimensional parameter space?
\end{itemize}

\subsection{Many-query scenario} \label{sec:mqRom}

We conclude the results demonstrating the application of
\rtworom~to a many-query analysis.
Using the same problem definition described in \S~\ref{sec:romAccuracy}, we assume
the period, $T$, parameterizing the forcing signal is
uncertain and follows a uniform distribution ${\mathcal U}(31, 69)$.
We aim to quantify how this parametric uncertainty impacts
the model predictions. This is a typical forward UQ study,
which we approach using Monte Carlo sampling.
%Of course, using the FOM to perform all the runs would be
%expensive, hence here we explore There are of course two ways we can do this: one would be to use the
%FOM and the other would be to construct and use the ROM to perform this study.
%Here, one key goal is to assess the computational gain resulting from
%using the ROM versus the FOM.

We use the \rtwofom~and \rtworom~to perform $P = 512$ realizations
of $T$ by randomly sampling ${\mathcal U}(31, 69)$.
For the ROM runs, we use $\romDim_{\vp}=436$ and $\romDim_{\stresses}=417$
(since these values proved to be accurate, see Figure~\ref{fig:romerrorsAcc2}(a)),
and select the most efficient combination of threads and $\nRuns$ by leveraging
the performance results shown in \S~\ref{sec:romScaling}.
Based on Figure~\ref{fig:romSpeedup}~(b) (since $\romDim=512$ is
a good approximation of $\romDim_{\vp}$ and $\romDim_{\stresses}$ used here),
we choose $\nRuns=64$ and $n=4$ threads to run each \rtworom{} simulation.
For the FOM, we use four threads $n=4$ and $\nRuns=4$ for each run.
This choice is made by using the FOM results
in \S~\ref{sec:fomScaling} to compute a metric similar to the one
computed for the ROM in \S~\ref{sec:romSpeedup}, and extracting
the most efficient combination. This is reported in
\S~SM2 of the supplemental material.

\subsubsection{Accuracy}
\begin{figure}[!t]
\centering
\begin{tikzpicture}
  \node[inner sep=0pt] (f00) at (0,0)  {\includegraphics[width=.49\textwidth]{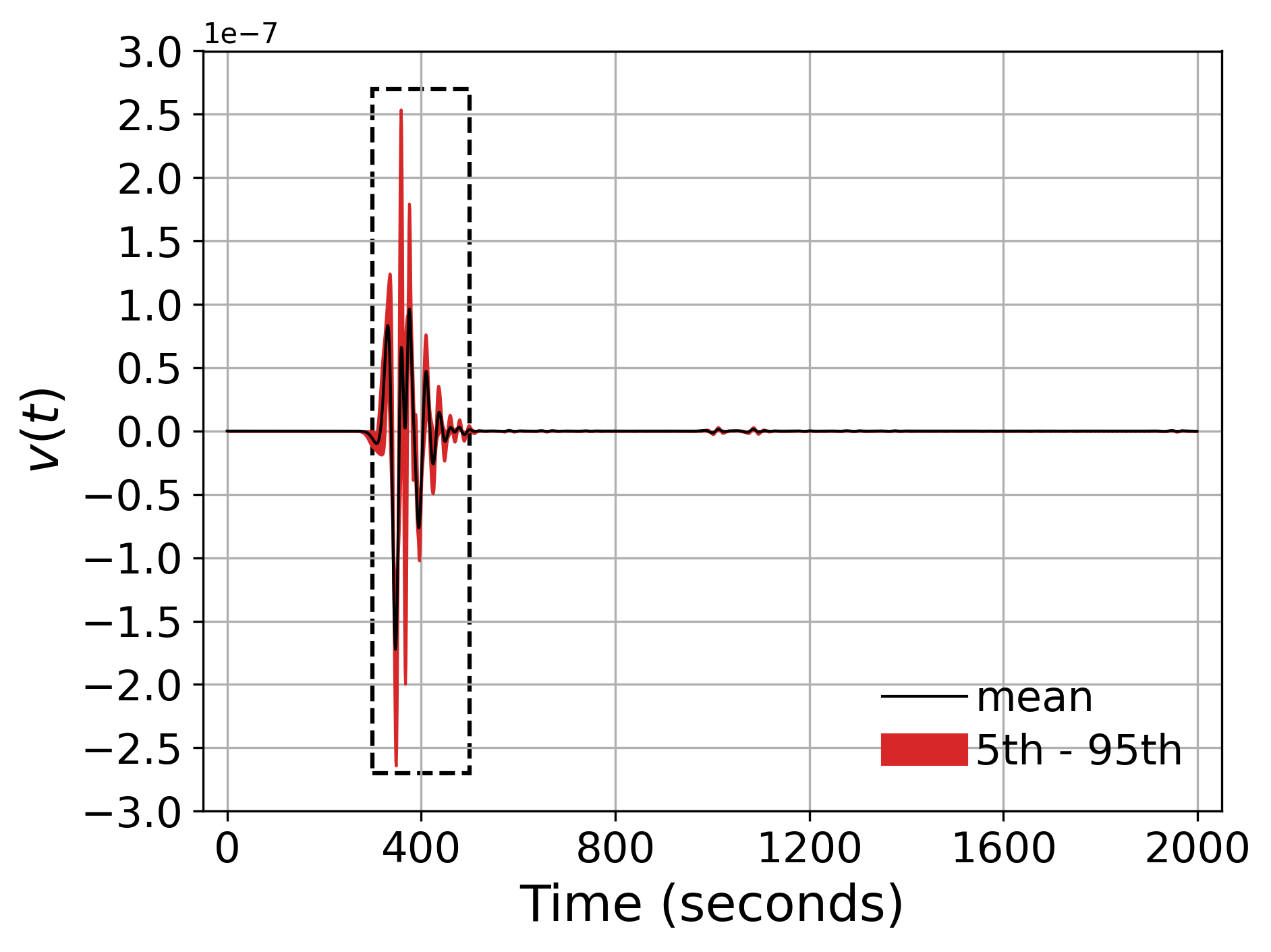}};
  \node[inner sep=0pt] (f01) at (6.3,0){\includegraphics[width=.49\textwidth]{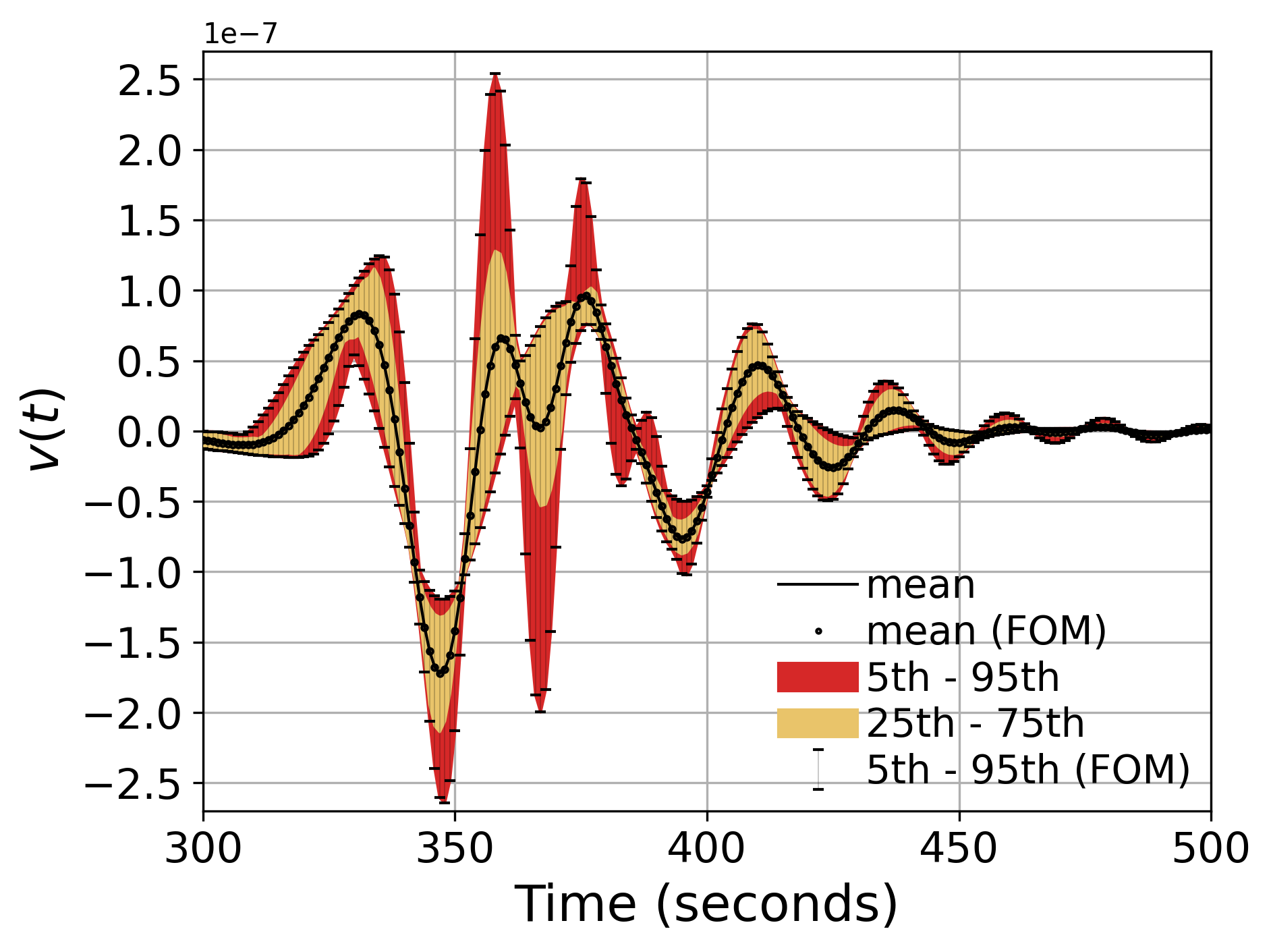}};
  \node[inner sep=0pt] (f00top) at (-1.27, 1.82){};
  \node[inner sep=0pt] (f00bot) at (-1.27, -1.4){};
  \node[inner sep=0pt] (f01top) at (4.2,  2.05){};
  \node[inner sep=0pt] (f01bot) at (4.2, -1.65){};
  \draw[-,thin] (f00top) -- (f01top);
  \draw[-,thin] (f00bot) -- (f01bot);
\end{tikzpicture}
\begin{tikzpicture}
  \node[inner sep=0pt] (f00) at (0,0)  {\includegraphics[width=.49\textwidth]{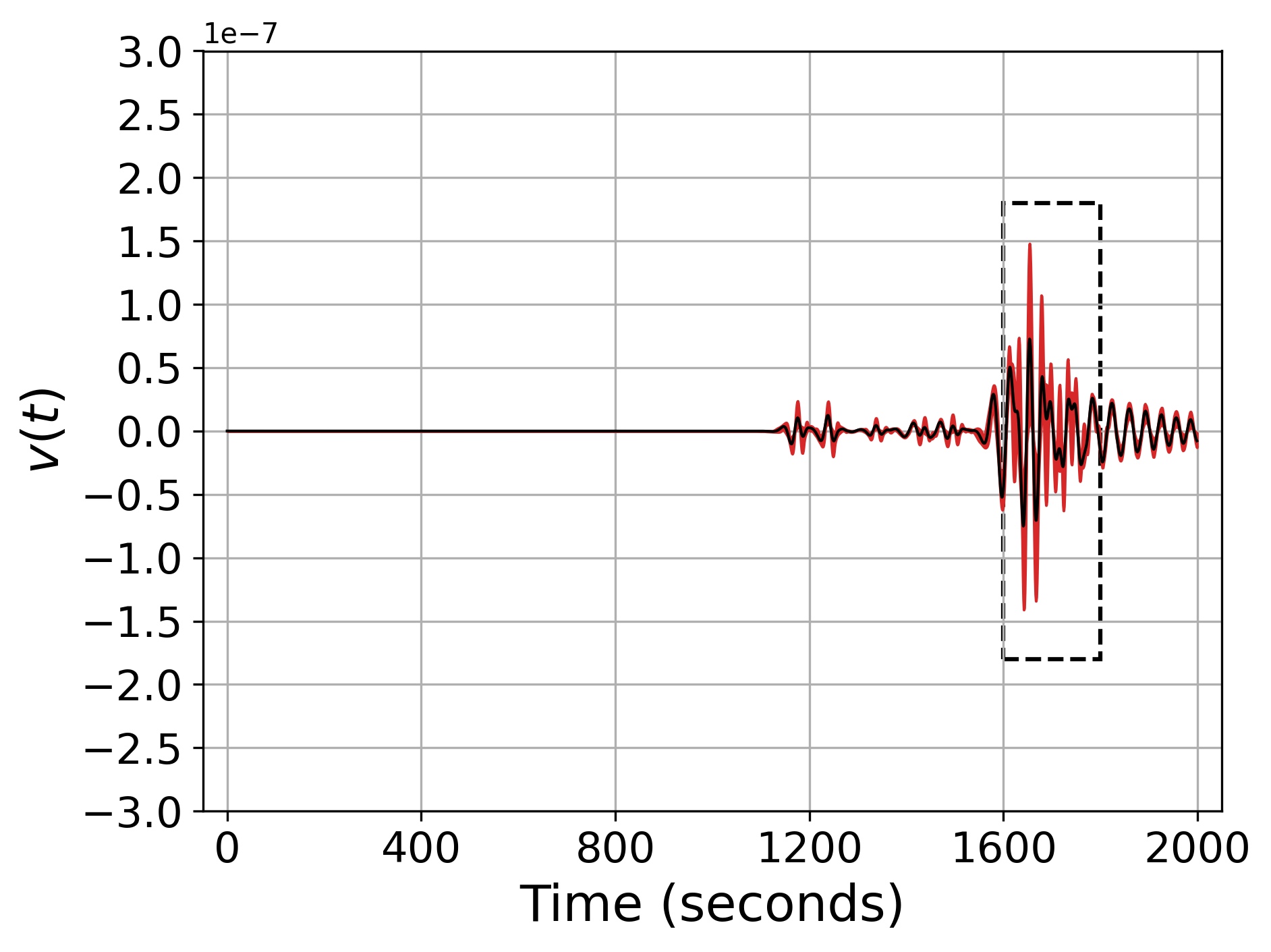}};
  \node[inner sep=0pt] (f01) at (6.3,0){\includegraphics[width=.49\textwidth]{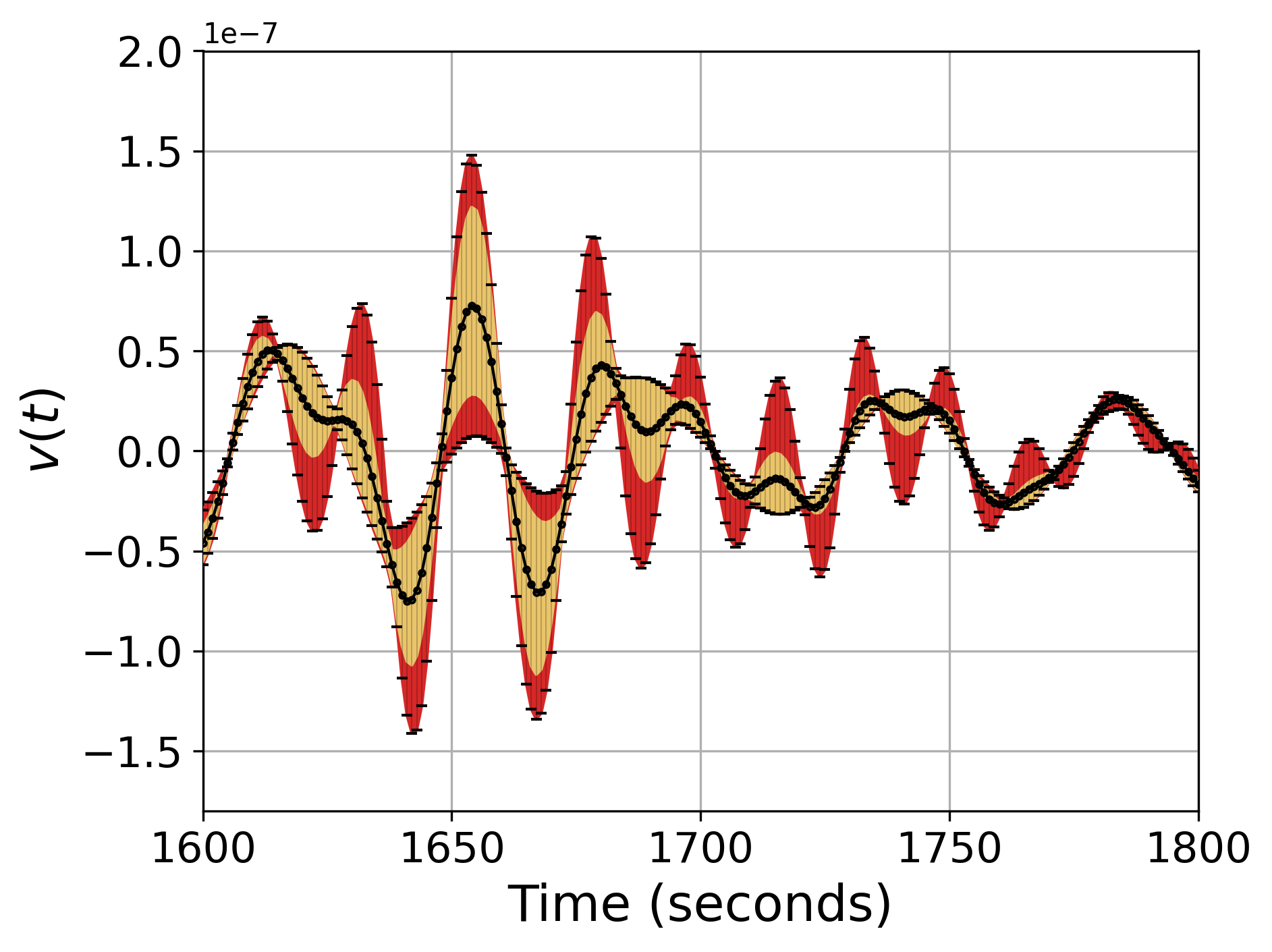}};
  \node[inner sep=0pt] (f00top) at (1.7, 1.29){};
  \node[inner sep=0pt] (f00bot) at (1.7, -0.85){};
  \node[inner sep=0pt] (f01top) at (4.2, 2.13){};
  \node[inner sep=0pt] (f01bot) at (4.2, -1.70){};
  \draw[-,thin] (f00top) -- (f01top);
  \draw[-,thin] (f00bot) -- (f01bot);
\end{tikzpicture}
\caption{Seismograms with representative statistics obtained
  at receivers on the earth surface, $r=6371$~km,
  with polar angles $\theta=10^\circ$~(first row), $60^\circ$~(second row),
  computed from the UQ analysis described in \S~\ref{sec:mqRom}.}
\label{fig:romuq}
\end{figure}
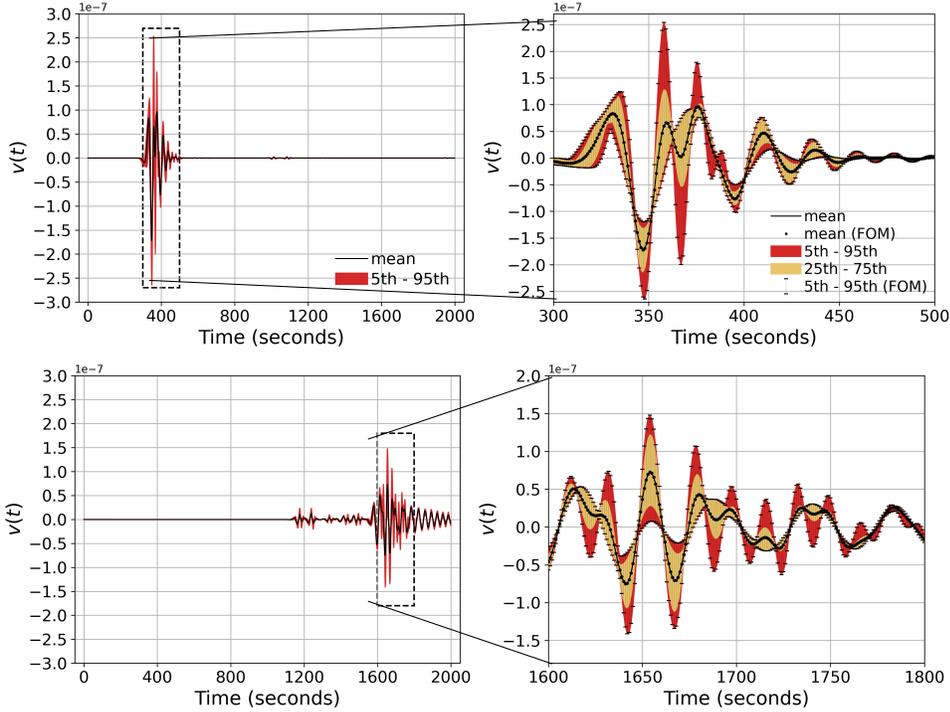

Figure~\ref{fig:romuq} shows the seismograms---and representative
statistics computed using the ensemble of 512 runs---collected
at locations on the earth surface, $r=6371$~km, with
polar angles $\theta=10^\circ$ and $60^\circ$.
The left column of Figure~\ref{fig:romuq} shows
the mean and the 5-95th percentile bounds over the full time domain
computed with the ROM using $\romDim_{\vp}=436$ and $\romDim_{\stresses}=417$.
The right column of Figure~\ref{fig:romuq} shows magnified views
of specific time windows displaying curves for different
percentiles as well as the mean and error bars computed with the FOM.
The predictions are characterized by substantial variations,
stemming from the complex pattern of reflections, refractions and interference
of the shear waves through the domain, and the ROM accurately
approximates the FOM results over the full time domain
in both its mean trend and statistics.
Note that in this case, the inherently complex dynamics of the waves
prevents one from being able to characterize the wavefield variability
just by evaluating the extreme values of the forcing period $T$.
It is therefore critical to sample the full range of $T$ to obtain
accurate statistics, which are fundamental to potentially
assess risk and extreme events probabilities.

\subsubsection{Computational Gain}
To quantify the computational gain, we can reason as follows.\\
{\it \rtworom{}.} Since we rely on a computing node with $36$ physical cores,
and use \rtworom{} with four threads and $\nRuns=64$,
we can launch 8 such simulations in parallel to compute $8 \times 64 = 512$ trajectories.
These 8 runs execute in parallel, and therefore finish in
approximately the same time as a single run with four threads and $\nRuns=64$,
which is about $7.5$ seconds for the current problem on the machine we used.\\
{\it \rtwofom{}.} For the FOM, we use individual runs with four threads and $\nRuns=4$,
implying that we can evaluate 8 runs in parallel, and complete
the full ensemble of $P$ trajectories with 16 sets of runs.
Each run takes about $455$ seconds, so the total runtime
is approximately $455 \times 16=7280$ seconds.
For the current problem and setup, \rtworom{}
is thus about $970$ times faster than the FOM.

If we were to run the same analysis using the \ronerom{},
we would launch 36 parallel single-threaded runs with $\nRuns=1$,
thus needing about $14$ sets of such runs to complete
the full ensemble of $P$ trajectories.
The approximate runtime of one such \ronerom{} is about $8$ seconds,
so the total runtime is approximately $8 \times 14=112$ seconds.
\rtworom{} is $15$ times faster than \ronerom{}.

Here we are not accounting for the cost
of computing the basis for the ROM simulations for two main reasons:
first, computing the POD basis only needs to be done once during the
offline stage; second, computing the POD for a very tall skinny matrix
(i.e., $m_r \gg m_c$ where $m_r$ and $m_c$ are the number of rows
and columns, respectively) can be efficiently done leveraging
one of the algorithms scaling with $m_c$, see,
e.g., chapter 45 in~\cite{Hogb06}.

We can summarize the following benefits in using the Galerkin ROM for this
seismic application: first, if needed, we can query the system very efficiently for
either a single or several values of $T$ to explore different statistics;
second, the dimensionality of the ROM is small enough such that one
can store the full time evolution of the generalized coordinates
and use them later on to efficiently reconstruct the full wave fields at target times;
third, computing the seismograms from the ROM data can be done very efficiently
because we can just use individual rows of the POD basis to reconstruct
the field at target points on the earth surface thus avoiding the need
to load the full basis matrix in memory.

\section{Conclusions} \label{sec:conclusions}

We presented a reformulation, called rank-2 Galerkin,
of the Galerkin ROM for LTI dynamical systems to convert them
from memory bandwidth- to compute-bound problems.
We described the details of the formulation and highlighted
its benefits compared to the formulation adopted so far in the community.
Rank-2 Galerkin is suitable to solve many-query problems
driven by the forcing term in a scalable and efficient way on modern computing node architectures.
%We also discussed its natural extension, called rank-3 Galerkin,
%which would allow one to also explore the parameters in the system matrix,
%and introduced the idea of exploiting the technological advancement
%of computing kernels used in deep learning, where tensors are key data structures.

We demonstrated the rank-2 Galerkin using, as a test case, the simulation
of elastic seismic shear waves in an axisymmetric domain.
We showed the excellent scalability of rank-2 Galerkin with
respect to the number of cores, which is a result of the nature
of the underlying computational kernels.
We quantified the accuracy of the ROM for this application as the period
of the forcing varies, and showed that for this problem, a Galerkin ROM
has excellent accuracy in both the reproductive and predictive scenarios.
Using a Monte Carlo sampling exercise, we demonstrated the
quantification of the uncertainty in the wavefields stemming from
a uncertain forcing period, and showed that rank-2 Galerkin is
about $970$ times faster than the FOM while being extremely accurate
in both the mean and statistics of the field.

As alluded to in this manuscript, several interesting future directions and applications arise.
In addition to problems of the form \eqref{eq:fom_ode}, 
the rank-2 formulation may yield computational gains in other ROM contexts, 
e.g., linearized systems (sensitivity and stability analysis), gradient-based solvers for nonlinear ROMs, 
and nonlinear ROMs with polynomial nonlinearities \cite{irinaSandReport}.
%For seismic reseach, one can compare different types of surrogates for the
%considered here, and extend to 3D seismic modeling.

The approach could be generalized to the case where the state and forcing are rank-3 tensors. 
This would enable one to {\it simultaneously} compute multiple realizations of $\paramsA$ (which parametrizes 
the system matrix) and $\paramsF$ (which parametrizes the forcing only). 
In this case, an obstacle to overcome would be to efficiently assemble 
the reduced system matrices for multiple realizations of $\paramsA$. 
The feasibility of this strictly depends on the parameterization of the system matrix: 
if the parameters can be decoupled from the matrix, then the reduced system matrices may be computed efficiently; 
if this decoupling is not possible, one could leverage interpolation methods 
applied to the system matrix \cite{amsallem2016}. 
This rank-3 formulation would open up interesting opportunities from a computational standpoint, 
for example employing higher-order singular value decompositions \cite{AFRA2016} and drawing ideas 
from the advances on tensor algebra taking place within the deep learning community, 
where rank-3 tensors are at the core of formulations. 
Of particular interest are algorithmic developments for 
batched matrix multiplication kernels \cite{shi2016, abdelfattah2016, li2019, lijuan2020} 
and hardware innovations such as tensor cores \cite{markidis2018} and tensor processing units \cite{jouppi2017}.

Beyond linear problems, one could assess the feasibility of applying a reformulation 
similar to the one presented here to ROMs of nonlinear systems, 
and the potential gains of applying the rank-2 or rank-3 Galerkin ROMs to inverse problems in geophysics.
Another critical aspect is the practical challenge related to 
implementing rank-2 Galerkin in a performance-portable way as a library to be used by other applications.
The latter topic is of particular interest to us: we are currently 
working on the design and implementation of the rank-2 Galerkin formulation 
in our ROM library Pressio\footnote{\url{https://github.com/Pressio}}.

%-------------------------------------------------------
\section*{Acknowledgments}
The authors thank Irina Tezaur for her helpful feedback.
This paper describes objective technical results and analysis.
Any subjective views or opinions that might be expressed in the paper
do not necessarily represent the
views of the U.S. Department of Energy or the United States Government.
This work was funded by the Advanced Simulation and Computing program
and the Laboratory Directed Research and Development program at
Sandia National Laboratories, a multimission laboratory managed and operated
by National Technology and Engineering Solutions of Sandia, LLC., a wholly
owned subsidiary of Honeywell International, Inc., for the U.S.\ Department of
Energy's National Nuclear Security Administration under contract
DE-NA-0003525.

%-------------------------------------------------------
%                        APPENDIX
%-------------------------------------------------------
\appendix

\section{Problem setup for the FOM scaling results} \label{sec:fomScalingSetup}
To run the FOM scaling tests presented in \S~4.1, we employ a material model comprising a
single, homogeneous layer with $\rho=2500$~(g/cm$^3$)
and ${v_s = 5000}$~(m/s). The source signal acts at $640$~(km) of depth
and takes the form of the first derivative of a Gaussian.
For the \ronefom{}, the source has a period $T=60$ (seconds)
and delay $180$~(seconds), while for the \rtwofom{} each forcing term
has the same Gaussian form but with a period
randomly sampled over the interval (60, 80).
The numerical dispersion condition is satisfied for all problems sizes.
The time step size is $\Delta t = 0.05$ such that numerical stability is ensured
for all problem sizes considered, and each run completes
$1000$ time steps, which allows us to collect meaningful
timing statistics of the computational kernels.

\section{When should an analyst prefer the Rank2FOM?} \label{sec:fomSpeedup}
In this section, we present for the FOM an analysis similar
to the one done for the ROM in \S~4.3.
Here, we set the constraints as follows.
%As a follow-up to the previous analysis, we now attempt the address the following problem:
%for a fixed problem size, we need to collect data by running $N$ (with $N$ large, e.g. $N>1e3$)
%samples of the focing term, as in a typical forward propagation study in UQ.
%We also need to meet some constraints.
First, we assume a limited budget of cores available, e.g.,
a single node with $36$ physical cores.
Second, we constrain the maximum number of forcing values running
simultaneously on the node to be $48$.
This can stem from, e.g., memory constraints.
Note that unlike the \rtworom{}, the \rtwofom{} cannot have too many simultaneous forcing terms being evaluated at the same
time on the node because the memory requirements are much bigger.
Therefore, this poses a much stricter limit, and we arbitrarily
choose $48$ as a constraint for the sake of the argument
noting, however, that it is quite a large value.
%% The question is: what combination of thread count and forcing size would
%% be most efficient to obtain those $N$ samples with those constraints?
%% This can be formulated as a discrete constrained optimization problem,
%% since we need to optimized over number of threads and forcing sizes,
%% and possibly over heterogenous setups with different runs using different
%% values of $f$ and different values of $n$.
%% Solving this problem is a general context is outside the scope of this work,
%% but here we provide some simple insights into this as follows.
%% The most basic scenario is one where we launch in parallel $18$ two-threaded runs,
%% each using $f=1$. This implies that all $36$ cores would be occupied,
%% and $18$ forcing realizations would be running at the same time, which means
%% that both the core budget and the memory constraint are satisfied.
%% A minor variation would be to have $9$ four-threaded runs at the same time each using $f=1$.
%% This would still satisfy both the core budget and the memory constraint.
%% The most interesting scenarios arise when we start changing the size of $f$.

Similarly to \S~4.3, we proceed as follows.
Let $\tau(\fomDim, n, \nRuns)$ represent the runtime to complete a {\it single}
FOM simulation of the form Eq.(3.2) with $\fomDim$ total degrees
of freedom using $n$ threads and a given value $\nRuns$.
It follows that the total runtime to complete trajectories for
$P$ forcing realizations with a budget of $36$ threads can be expressed as
\begin{equation}
\tau^{P}(\fomDim, n, \nRuns) = \tau(\fomDim, n, \nRuns) \frac{P}{\frac{36}{n} \nRuns},
\end{equation}
because $\frac{36}{n}$ is the number of independent runs executing
in parallel on the node with each run responsible of computing $\nRuns$ trajectories.
We can define the following metric
\begin{equation} \label{eq:fomSpeedup}
  s(\fomDim,n,\nRuns)
  = \frac{\tau^P(\fomDim,2,1)}{\tau^P(\fomDim, n, \nRuns)}
  = \frac{\tau(\fomDim,2,1)}{\tau(\fomDim, n, \nRuns)} \frac{2\nRuns}{n},
\end{equation}
where $s(\fomDim,n,\nRuns)>1$ indicates \rtwofom~is more efficient
than \ronefom, while the opposite is true for $s(\fomDim,n,\nRuns)<1$.
We remark that the reference case used here for this metric is
the \ronefom{} using $n=2$ threads, since for the FOM we
never use less than $2$ threads, see \S~4.1.
\begin{figure}[t]
  \centering
  \includegraphics[trim=0 0 0 0,clip,width=0.47\textwidth]
                  {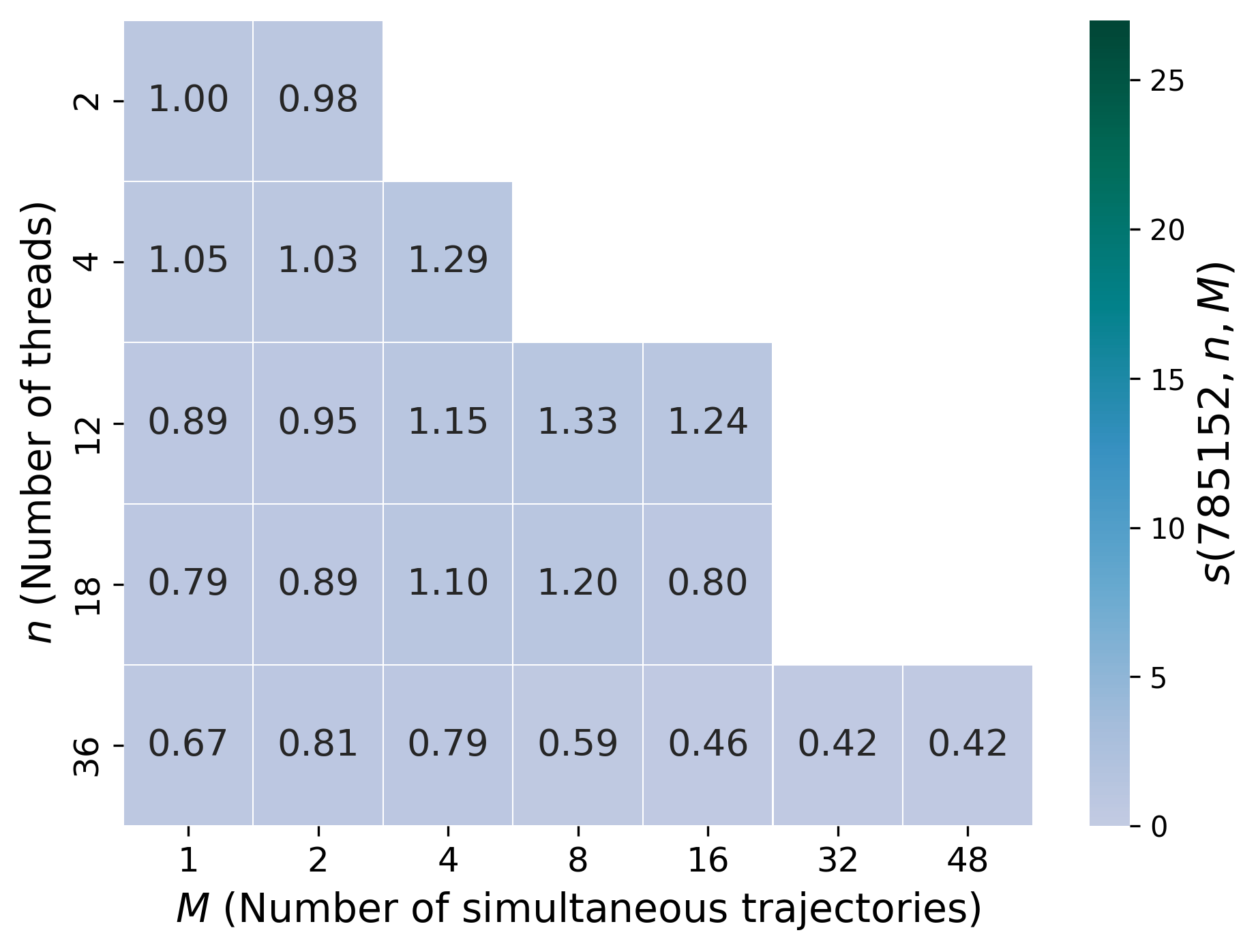}(a)
  \includegraphics[trim=45 0 0 0,clip,width=0.43\textwidth]
                  {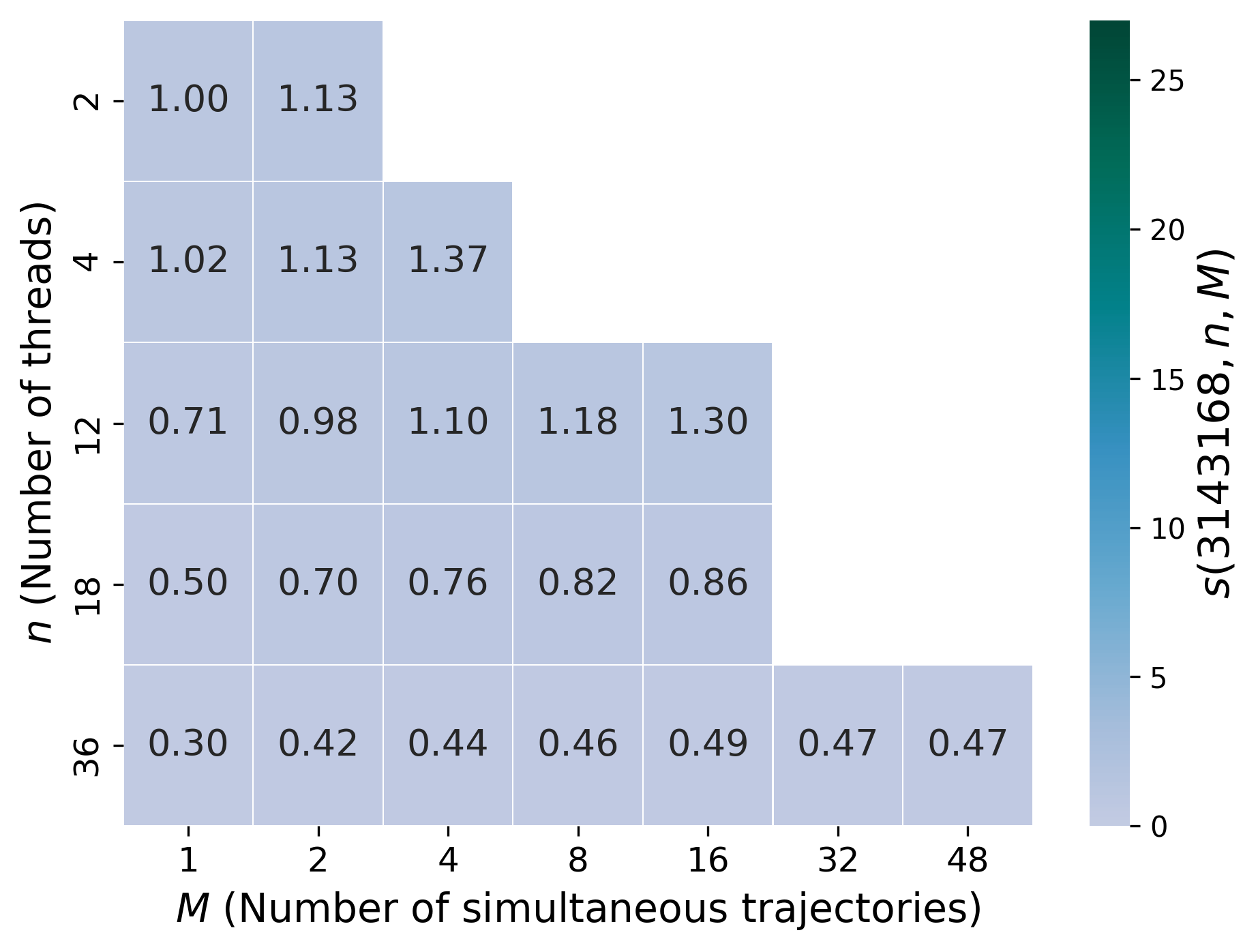}(b)\\
  \centering
  \includegraphics[trim=0 0 0 0,clip,width=0.47\textwidth]
                  {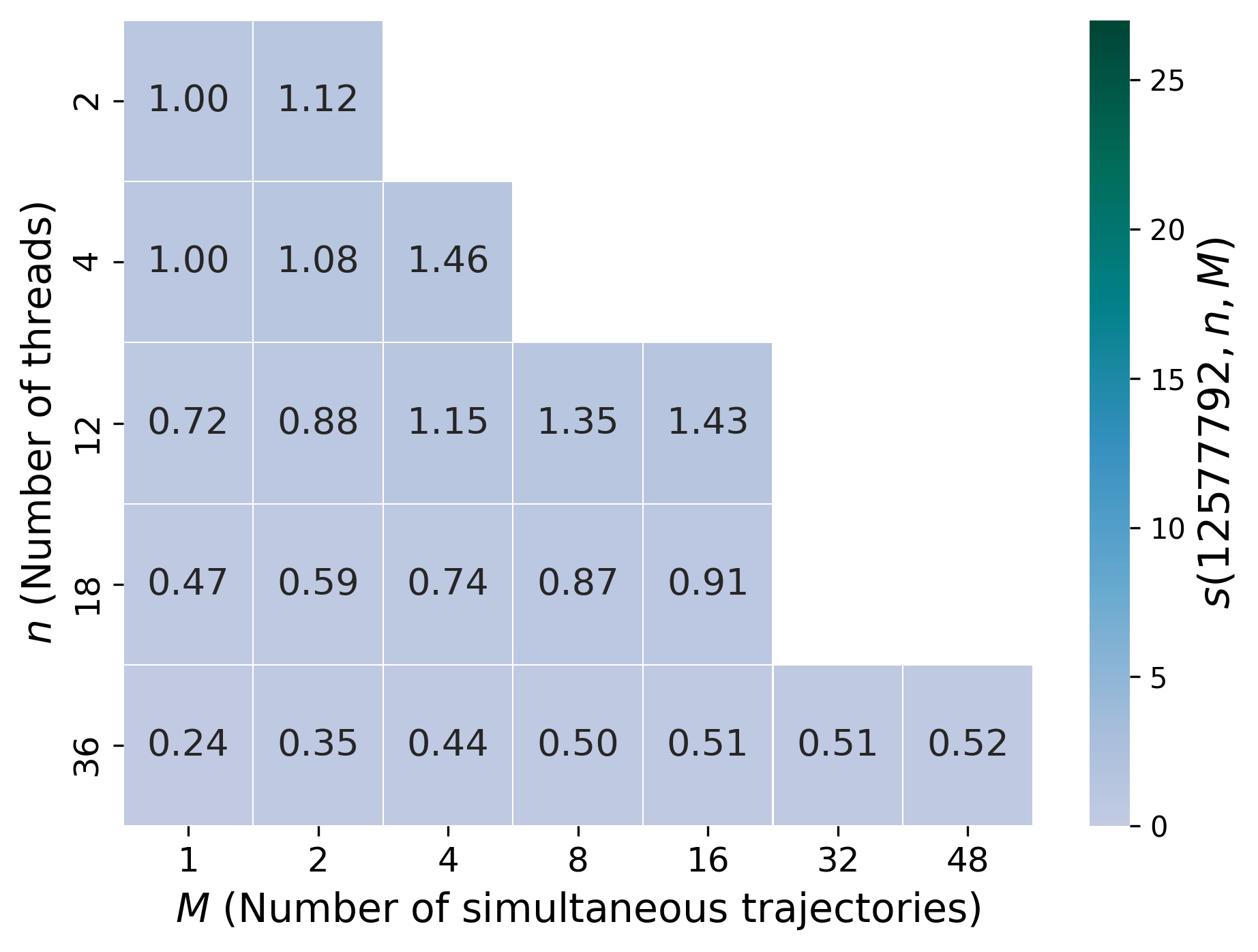}(c)
  \includegraphics[trim=45 0 0 0,clip,width=0.43\textwidth]
                  {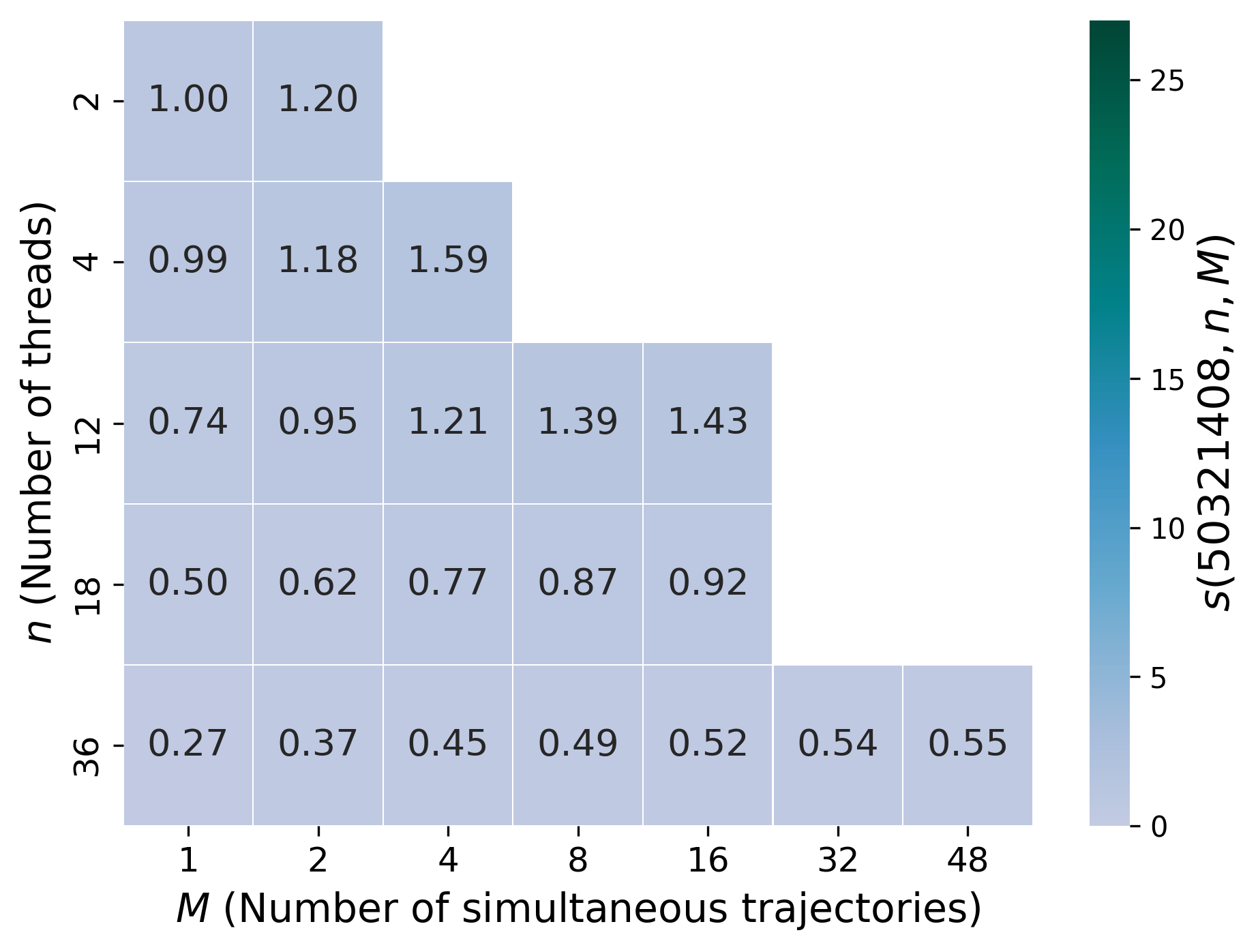}(d)
  \caption{Heatmap visualization of $s(\fomDim, n, \nRuns)$ (see Eq.~\ref{eq:fomSpeedup})
    computed for $\fomDim = 785,152$~(a), $3,143,168$~(b), $12,577,792$~(c) and $50,321,408$~(c),
    and various values of $\nRuns$ and $n$. The white region in each plot identifies
    the cases violating the constraints listed in \S~\ref{sec:fomSpeedup}.
    Note that we set the colorbar with the same limits used in \S~4.3 to facilitate
    the comparison between the two sets of results.}
\label{fig:fomSpeedup}
\end{figure}

Figure~\ref{fig:fomSpeedup} shows a heatmap visualization of
$s(\fomDim, n, \nRuns)$ computed for various values of $\nRuns$
and $n$, and $\fomDim \in \{785,152; 3,143,168; 12,577,792; 50,321,408\}$.
To generate these plots, we used the FOM runtimes obtained in \S~4.1.
The heatmap plot allows us to reason as follows.
For example, for the $3 \cdot 10^6$ problem size, using $\nRuns=4$
and $n=4$ is $37\%$ more efficient than the case $\nRuns=1,n=2$,
and becomes $59\%$ more efficient if we consider the $50 \cdot 10^6$ problem size.
For $\nRuns=16$ and $n=12$, the gain is about $30\%$ for the $3 \cdot 10^6$
case and becomes $43\%$ for the $50 \cdot 10^6$ one.
This trend suggests that the gain increases with the problem size.
%It is interesting to observe that the cases
%yielding a speedup %with respect to $\nRuns=1,n=2$
%are the same regardless of the
%problem size, and only the magnitude of the gain changes.
It is interesting to note that the gains are overall quite small,
never exceeding $60\%$. This is due to the fact that
while the \rtwofom{} has a higher compute intensity than \ronefom{},
they are both memory-bandwidth bound problems, and therefore they are
both limited. This is in contrast to what we observed for the ROM in \S~4.3,
where we obtained up to 26 times speedup using the rank-2 ROM formulation.

\section{Data and Code Availability}
The code developed for this work and data used in the paper are publicly available at:
\begin{itemize}
\item Code repository:~\hspace{.2cm}\url{https://github.com/fnrizzi/ElasticShearWaves}\\
  Git hash: 2e4cf2151d3e284b4ce3c29f7f6a12efa097915b

\item The full dataset:\hspace{.3cm}\url{https://github.com/fnrizzi/RomLTIData}
  contains the data as well as the scripts to generate the plots
\end{itemize}

%-------------------------------------------------------
%                        BIBLIO
%-------------------------------------------------------
\bibliographystyle{siamplain}
\bibliography{references}
\end{document}